\documentclass[
reprint, longbibliography,
superscriptaddress,
preprintnumbers,nofootinbib,
nobibnotes,
amsmath,amssymb,
aps, pra,
floatfix,
]{revtex4-2}

\usepackage[normalem]{ulem}
\usepackage{gensymb}
\usepackage{upgreek}
\usepackage{mathrsfs}
\usepackage{comment}
\usepackage{graphicx}
\usepackage{dcolumn}
\usepackage{bm}
\usepackage{float}
\usepackage{url}
\usepackage[]{hyperref}
\usepackage{xcolor}
\usepackage{textcomp}
\usepackage{amsmath}
\usepackage{varwidth}
\usepackage{placeins}
\usepackage{gensymb}

\usepackage{siunitx}
\hypersetup{
    colorlinks,
    linkcolor={red!50!black},
    citecolor={blue!50!black},
    urlcolor={blue!80!black}
}
\usepackage[center]{titlesec}
  \titleformat{\section}
   {\centering\normalfont\fontsize{10}{10}\bfseries}{\thesection}{1em}{}
  \titlespacing{\section}{0pt}{12pt plus 4pt minus 2pt}{6pt plus 2pt minus 2pt}
    \titleformat{\subsection}
   {\centering\normalfont\fontsize{10}{10}\bfseries}{\thesubsection}{1em}{}
  \titlespacing{\subsection}{0pt}{12pt plus 4pt minus 2pt}{6pt plus 2pt minus 2pt}

\usepackage{tabularx}
\newcolumntype{C}{>{\centering\arraybackslash}X} 
\usepackage{multirow}

\usepackage{amsfonts, amsmath, amssymb, graphicx, xcolor, verbatim, lipsum, dsfont, bm, hyperref, makecell, comment, upgreek, graphicx, dcolumn, array}

\usepackage[export]{adjustbox}
\usepackage[mathlines]{lineno}
\usepackage[T1]{fontenc}

\DeclareLanguageAlias{en}{english}
\DeclareLanguageAlias{eng}{english}

\begin{document}

\title{Chemically resolved nuclear magnetic resonance spectroscopy by longitudinal magnetization detection with a diamond magnetometer}

\author{Janis~Smits}
\email[Email address: ]{smitsjanis@gmail.com}
\affiliation{Center for High Technology Materials, 
University of New Mexico, Albuquerque, NM, USA}

\author{Yaser~Silani}
\affiliation{Center for High Technology Materials, 
University of New Mexico, Albuquerque, NM, USA}

\author{Zaili~Peng}
\affiliation{Center for High Technology Materials, 
University of New Mexico, Albuquerque, NM, USA}

\author{Bryan~A.~Richards}
\affiliation{Center for High Technology Materials, 
University of New Mexico, Albuquerque, NM, USA}
\affiliation{Department of Physics and Astronomy,
University of New Mexico, Albuquerque, NM, USA}

\author{Andrew~F.~McDowell}
\affiliation{NuevoMR, Albuquerque, NM, USA}

\author{Joshua~T.~Damron}
\affiliation{Center for High Technology Materials, 
University of New Mexico, Albuquerque, NM, USA}
\affiliation{Oak Ridge National Laboratory, Oak Ridge, TN, USA}

\author{Maxwell~D.~Aiello}
\affiliation{Center for High Technology Materials, 
University of New Mexico, Albuquerque, NM, USA}
\affiliation{Department of Physics and Astronomy,
University of New Mexico, Albuquerque, NM, USA}

\author{Maziar~Saleh~Ziabari}
\affiliation{Center for High Technology Materials, 
University of New Mexico, Albuquerque, NM, USA}
\affiliation{Department of Physics and Astronomy,
University of New Mexico, Albuquerque, NM, USA}

\author{Andrey~Jarmola}
\email[Email address: ]{andrey.jarmola@gmail.com}
\affiliation{ODMR Technologies Inc., El Cerrito, CA, USA}
\affiliation{Department of Physics, University of California, Berkeley, CA, USA}

\author{Victor~M.~Acosta}
\email[Email address: ]{victormarcelacosta@gmail.com}
\affiliation{Center for High Technology Materials, 
University of New Mexico, Albuquerque, NM, USA}
\affiliation{Department of Physics and Astronomy,
University of New Mexico, Albuquerque, NM, USA}

\date{\today}

\begin{abstract}
Non-inductive magnetometers based on solid-state spins offer a promising solution for small-volume nuclear magnetic resonance (NMR) detection. A remaining challenge is to operate at a sufficiently high magnetic field to resolve chemical shifts at the part-per-billion level. Here, we demonstrate a Ramsey-$M_z$ protocol that uses Ramsey interferometry to convert an analyte's transverse spin precession into a longitudinal magnetization ($M_z$), which is subsequently modulated and detected with a diamond magnetometer. We record NMR spectra at $B_0=0.32~{\rm T}$ with a fractional spectral resolution of ${\sim}350~{\rm ppb}$, limited by the stability of the electromagnet bias field. We perform NMR spectroscopy on a ${\sim}1~{\rm nL}$ detection volume of ethanol and resolve the chemical shift structure with negligible distortion. Through simulation, we show that the protocol can be extended to fields up to $B_0=3~{\rm T}$, with minimal spectral distortion, using composite nuclear-spin inversion pulses. For sub-nanoliter analyte volumes, we estimate a resolution of ${\sim}1~{\rm ppb}$ and concentration sensitivity of ${\sim}40~{\rm mM \,s^{1/2}}$ is feasible with improvements to the sensor design. Our results establish diamond magnetometers as high-resolution NMR detectors in the moderate magnetic field regime, with potential applications in metabolomics and pharmaceutical research.
\end{abstract}

\maketitle

\section{\label{sec:level1}Introduction}
Nuclear magnetic resonance (NMR) spectroscopy is a powerful analytical technique used widely in chemistry, biology, and pharmaceutical metabolomics research~\cite{BEC1993,BOE2024,MAR2017}
, but it is often hindered by poor signal-to-noise ratio (SNR), particularly for small sample volumes. In conventional, inductive NMR detection, the SNR depends on the external field strength, scaling approximately as $B_0^{7/4}$~\cite{HOULT1976}. The chemical specificity also improves with increasing $B_0$, since spectral splittings due to chemical shifts increase proportional to $B_0$. This has motivated the development of increasingly large superconducting magnets, but progress has plateaued in recent decades~\cite{ARD2015}. Miniature inductive coils have been developed to improve SNR for small-volume samples~\cite{STO1997,MCD2007,FUG2017,GRI2017}, but their inherent discontinuities in magnetic susceptibility can lead to inadequate spectral resolution for demanding applications like microfluidic metabolomic analysis~\cite{BRK2011}. 

An alternative strategy for small-volume NMR spectroscopy is to use a non-inductive magnetometer for detection. Various sensors have been pursued, including those based on superconducting quantum interference devices~\cite{SET1997,MCD2002}, magnetoresistance sensors~\cite{VER2008,OOG2021}, alkali-metal vapor magnetometers~\cite{SAV2005,LED2008}, and, recently, magnetometers based on nitrogen-vacancy (NV) centers in diamond~\cite{MAM2013,STA2013,ASL2017,GLE2018,SMI2019,REN2023,BRU2023}. When using diamond magnetometers to detect thermally polarized nuclei (NV NMR), the SNR can be superior to inductive detection at lower frequencies~\cite{SIL2023}, and it scales linearly with $B_0$, supporting the intriguing possibility of working at moderate magnetic fields, $B_0\lesssim3~{\rm T}$, where compact permanent magnets are available. Moreover, while the sensitivity of inductive NMR detectors is fundamentally limited by thermal Johnson noise~\cite{HOU1976}, diamond magnetometers are limited by spin-projection and photon shot noise~\cite{TAY2008}, which is predicted to lead to better SNR at moderate fields~\cite{GLE2018,SIL2023}. Finally, diamond membranes feature a planar sample-sensor interface, which may reduce magnetic field gradients compared to microcoil geometries, potentially enabling picoliter-scale NMR spectroscopy with part-per-billion (ppb) spectral resolution~\cite{SMI2019}.

\begin{figure*}[htb]
    \centering
    \includegraphics[width=0.98\textwidth]{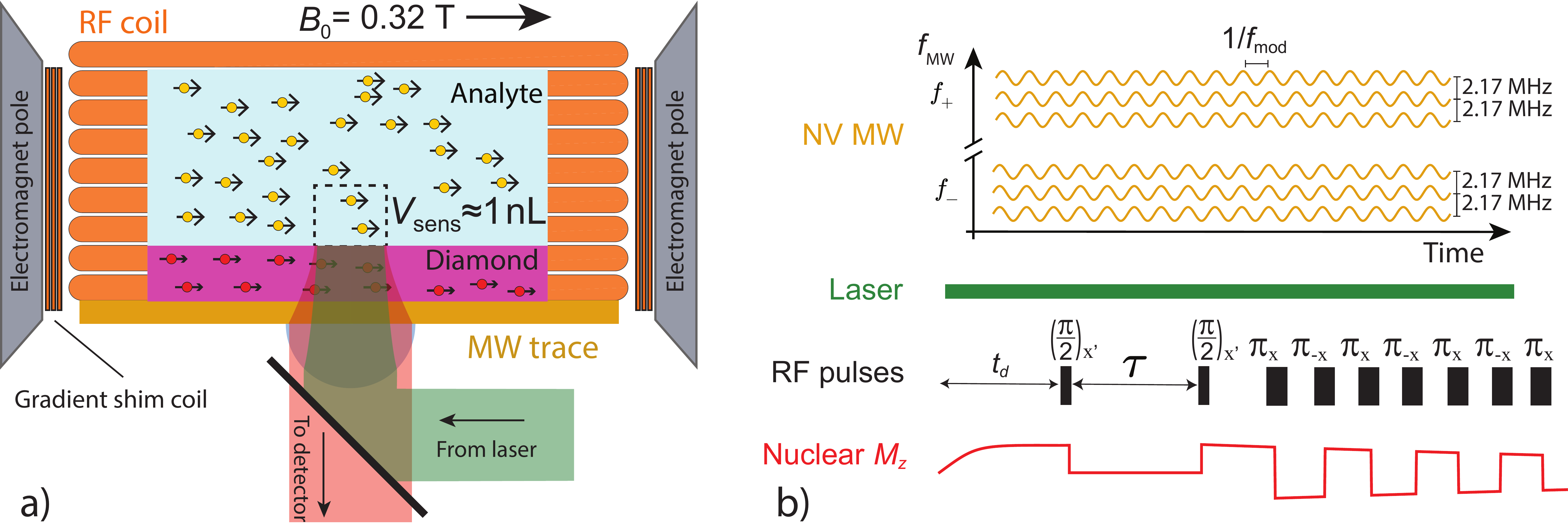}
    \caption{\textbf{Ramsey-$\bm{M_z}$ NV NMR.} (a) Side view depiction of the apparatus. A magnetic field $B_0=0.32~{\rm T}$ is applied along the NV axis using a shimmed electromagnet. Liquid analyte is in direct contact with a diamond adhered to a glass slide containing a copper microwave (MW) trace. NV laser excitation and fluorescence collection are done through a window between microwave traces. A resonant RF coil is wound around the analyte container to drive proton spins and monitor the conventional NMR signal. The effective analyte detection volume is $V_{\rm sens}\approx 1~{\rm nL}$ (\ref{sec:SI_nucmag}). (b) Ramsey-$M_z$ timing diagram. Microwave tones drive all six NV $f_{\pm,i}$ transitions, Eq.~\eqref{eq:fpm}, simultaneously. Each tone is frequency modulated at a rate $f_{\rm mod}=10.1~{\rm kHz}$ to generate an AC fluorescence signal for lock-in detection. Laser illumination and DNP microwaves (if used) are applied continuously. Phase-cycled RF pulses (${\sim}13.8~{\rm MHz}$, $20~{\rm \upmu s}$ duration $\pi$ pulses, $10~{\rm \upmu s}$ $\pi/2$ pulses) provide a modulated nuclear $M_z$ signal, with an amplitude that depends on the NMR frequency, Eq.~\eqref{eq:magfield_nv}, that is detected by the diamond magnetometer.
    }
    \label{fig:fig1}
\end{figure*}

Most NV NMR experiments to date operated at magnetic fields, $B_0<0.2~{\rm T}$, where it is convenient to detect oscillating \textit{transverse} nuclear magnetization. Typically, a synchronized AC magnetometry protocol is applied, consisting of a train of NV electron-spin microwave $\pi$ pulses spaced $\tau_{\rm rep}\approx1/(2f_p)$ apart, where $f_p$ is the NMR frequency~\cite{GLE2018,SMI2019,BRU2023}. These experiments realize remarkable absolute spectral resolution, down to ${\sim}0.65~{\rm Hz}$~\cite{SMI2019}, but the ${\gtrsim}1~{\rm ppm}$ fractional resolution is insufficient to discern chemical shifts at the level needed for routine solution-state NMR spectroscopy. In typical liquid samples, the natural NMR linewidth of protons is ${\sim}0.1~{\rm Hz}$~\cite{HAR2010}, so a magnetic field $B_0\gtrsim0.3~{\rm T}$ is needed to realize sub-$10~{\rm ppb}$ resolution. At these fields, AC detection of transverse nuclear magnetization is impractical. For example, at $B_0=1.4~{\rm T}$, the proton NMR frequency is $f_p=60~{\rm MHz}$, and the required $\pi$-pulse repetition interval is $\tau_{\rm rep}\approx8~{\rm ns}$. Considering that the NV transition frequencies are fairly high at this field, $f_{\pm} \approx 40~{\rm GHz}$, this is a severe technical challenge. Evidently, a new approach is needed, and several recent proposals have attempted to address this~\cite{CAS2018,MUN2020,MUN2023,MEI2023,DAL2024}.

Here, we demonstrate high-resolution NV NMR spectroscopy by detecting \textit{longitudinal} nuclear magnetization. Similar strategies have been implemented with other NMR detectors, including Jospehson junctions~\cite{DAY1972}, cantilevers~\cite{MAD2004,POG2010}, magnetoresistance sensors~\cite{VER2008}, atomic magnetometers~\cite{MOU2023}, and even single NV centers~\cite{MAM2013}, but (until now) not in high-resolution NV NMR spectroscopy. Our ``Ramsey-$M_z$'' protocol uses Ramsey interferometry to encode the phase accumulated by analyte nuclear magnetization during a free precession interval into a longitudinal magnetization amplitude. The longitudinal magnetization is then converted to a square-wave AC magnetic field, using a train of nuclear $\pi$ pulses, and concurrently detected with a broadband diamond magnetometer. With this technique, we record NMR spectra at $B_0=0.32~{\rm T}$ with a fractional spectral resolution of ${\sim}350~{\rm ppb}$, limited by the stability of the electromagnet field. We perform NMR spectroscopy on a ${\sim}1~{\rm nL}$ detection volume of ethanol and resolve the chemical shift structure with negligible distortion. Finally, we extrapolate the performance of the Ramsey-$M_z$ protocol up to $B_0=3~{\rm T}$ and describe modifications that would result in a sub-nL concentration sensitivity two orders of magnitude better than that of inductive detection.

\section{Experimental setup}
The experimental apparatus is depicted in Fig.~\ref{fig:fig1}(a).  Additional details are provided in \ref{sec:SI_exp_layout}. When detecting longitudinal nuclear magnetization ($M_z$), the optimal direction of the magnetic field with respect to the sample-sensor interface is different than that for detecting transverse magnetization ($M_{\perp}$)~\cite{GLE2018,SMI2019}. In both cases the magnetic field is aligned along one of the four NV crystallographic axes. However, in $M_z$ detection, the NV axes should ideally lie either in the plane of the sample-sensor interface or normal to it~(\ref{SI:diamond_cut}).  

In our experiment, the $B_0$ bias magnetic field is applied by an electromagnet with first- and second-order gradient shim coils~\cite{SMI2019}. We use a [110]-cut diamond membrane and position it such that $B_0$ is aligned along one of the in-plane NV axes. The membrane is formed from a $^{12}$C-enriched diamond, with an NV concentration of ${\sim}4~{\rm ppm}$, that is cut and polished to dimensions ${\sim}250{\times}250{\times}60~{\rm \upmu m^3}$. The diamond membrane is adhered to a glass slide that has two copper microwave traces; one is used to drive NV spin transitions, and the other is sometimes used to facilitate Overhauser dynamic nuclear polarization (DNP) transfer from TEMPOL radicals~\cite{RAV2016, LEE2019, MAL2021, BUC2020} to nuclear spins within the analyte. A 3D-printed photopolymer container, holding ${\gtrsim}1~{\rm \upmu L}$ of liquid analyte, is adhered to the glass slide such that the analyte is in direct contact with the diamond. A resonant radio-frequency (RF) coil is wound around the container to drive analyte nuclear spin transitions. 

\begin{figure*}[htb]
   \centering
    \includegraphics[width=0.95\textwidth]{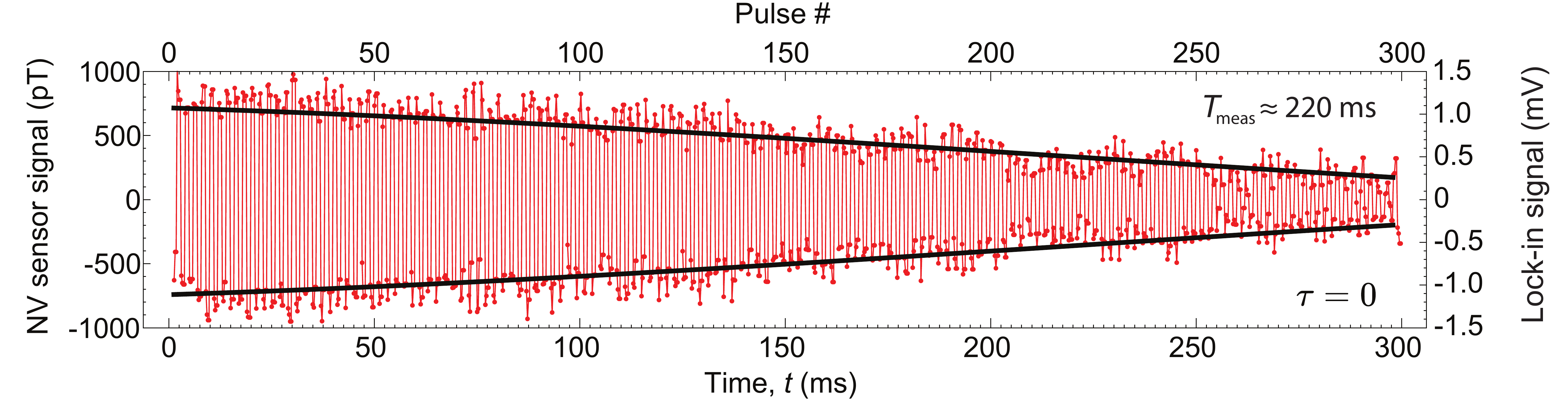}
    \caption{\textbf{Modulated $\bm{M_z}$ signal.} Diamond magnetometer signal, $B_{\rm nuc}(t)$, of water for zero Ramsey delay ($\tau = 0$). The lock-in output voltage is converted to picotesla using independent calibrations (\ref{SI:mag_calib}). Phase-cycled RF $\pi$ pulses are applied every $1~{\rm ms}$. Each data point is the average over a $0.25~{\rm ms}$ interval. Intervals containing RF pulses are omitted. A Guassian fit to the signal envelope reveals a decay time, $T_{\rm meas}\approx220~{\rm ms}$}
    \label{fig:fig1p5}
\end{figure*}

NV centers are optically excited by ${\sim}80~{\rm mW}$ of $532~{\rm nm}$ laser light that is directed through a window between the microwave traces. A high-refractive-index half-ball lens, epoxied to the glass slide, and an aspheric lens are used to collect and collimate ${\sim}0.1~{\rm mW}$ of NV fluorescence, which then passes through a dichroic mirror and is imaged onto a photodetector (\ref{sec:SI_exp_layout}). The effective sensor volume is defined by the product of the excitation beam area, ${\sim}\pi/4\cdot(135~{\rm \upmu m})^2$, and diamond membrane thickness, ${\sim}60~{\rm \upmu m}$. Magnetostatic calculations (\ref{sec:SI_sens_vol}) indicate that ${\gtrsim}50\%$ of the $M_z$ NMR signal comes from a $1~{\rm nL}$ cylinder of analyte centered about the excitation beam. 

A schematic of the detection protocol is shown in Fig.~\ref{fig:fig1}(b). A broadband diamond magnetometer continuously records the local magnetic field via a continuous-wave optically-detected magnetic resonance (CW-ODMR) protocol. The NV center ground state is comprised of two unpaired electrons, with total spin $S=1$, coupled to a $I=1$ $^{14}$N nuclear spin. For a magnetic field $B_z$ aligned along the NV axis, the allowed electron spin transitions are given by:
\begin{equation}
\label{eq:fpm}
f_{\pm,i} = |D \pm (\gamma_{\rm nv} B_z + A_{||} m_i)|, 
\end{equation}
where $D\approx2.87~{\rm GHz}$ is the zero-field splitting, $\gamma_{\rm nv}=28.03~{\rm GHz/T}$ is the NV gyromagnetic ratio, $A_{||}=-2.17~{\rm MHz}$ is the axial hyperfine coupling, and $m_i$ denotes the $^{14}$N nuclear spin sub-level. 

As shown in Eq.~\eqref{eq:fpm}, there are six allowed spin transitions, and maximizing the CW-ODMR signal contrast calls for addressing all of them simultaneously. To do so, while minimizing the effect of the temperature dependence of $D$, we apply a dual-resonance protocol~\cite{WOJ2018,FES2020}. Microwave tones are applied at all six $f_{\pm,i}$ resonance frequencies, and each tone is frequency-modulated at a $10.1~{\rm kHz}$ rate and ${\sim}500~{\rm kHz}$ deviation (\ref{sec:SI_MW_block}). Since $\gamma_{nv} B_0>D$, we choose the phase of the frequency-modulation waveform to be the same for all six tones, which minimizes dependence on temperature while maximizing sensitivity to changes in $B_z$ (\ref{sec:SI_MW_block}). The resulting fluorescence photodetector signal is demodulated with a lock-in amplifier and used to measure the magnetic field produced by the analyte $M_z$. The demodulated signal is also used in a low-frequency feedback loop to drive a secondary set of trim coils that compensates for environmental field drift (\ref{SI:mag_stab}). This feedback system allows for a $B_0$ field stability of $\lesssim350~{\rm ppb}$ over the course of several hours. 

An additional coil is used to provide independently-calibrated test signals that are used to convert the demodulated signal into magnetic field units~(\ref{SI:mag_calib}). The experimentally-determined sensitivity of the diamond magnetometer is ${\sim} 100~{\rm pT_{rms}\,s^{1/2}}$ for magnetic fields oscillating at approximately $100\mbox{-}1000~{\rm Hz}$. The observed noise floor is mostly accounted for by the expected contributions due to photon shot noise~\cite{TAY2008} and microwave phase noise~\cite{BER2024} (\ref{sec:SI_sens}).

In the Ramsey-$M_z$ protocol, two phase-coherent radio-frequency (RF) $\pi/2$ pulses, spaced by a variable time $\tau$, are applied at a frequency detuning $\Delta_{\rm nuc}$ with respect to the proton NMR frequency. The resulting $M_z$ component of the nuclear-spin magnetization is given by:
\begin{equation}
    M_z(\tau) = M_0 \cos(2\pi\Delta_{\rm nuc}\tau)\, e^{-\tau/T_2^{\ast}},
\end{equation}
where $M_0$ is the equilibrium magnetization magnitude prior to the Ramsey sequence and $T_2^{\ast}$ is the analyte nuclear-spin dephasing time. Next, a train of resonant RF $\pi$ pulses modulates the sign of $M_z$. This effectively mixes the slowly-decaying $M_z$ signal to an AC signal at a frequency (typically ${\sim}500~{\rm Hz}$) where the diamond magnetometer sensitivity is optimized. The phase of each RF $\pi$ pulse is alternated by $180\degree$ to reduce the impact of pulse imperfections \cite{MAD2004}. Finally, a dead time $t_d=0.6~{\rm s}\approx T_1$, where $T_1$ is the nuclear longitudinal spin relaxation time, is imposed to allow analyte magnetization to approach equilibrium. This sequence is repeated for different values of $\tau$ to obtain a Ramsey interferogram analogous to a free-induction decay in traditional NMR spectroscopy.

\begin{figure*}[t]
    \centering
    \includegraphics[width=0.75\linewidth]{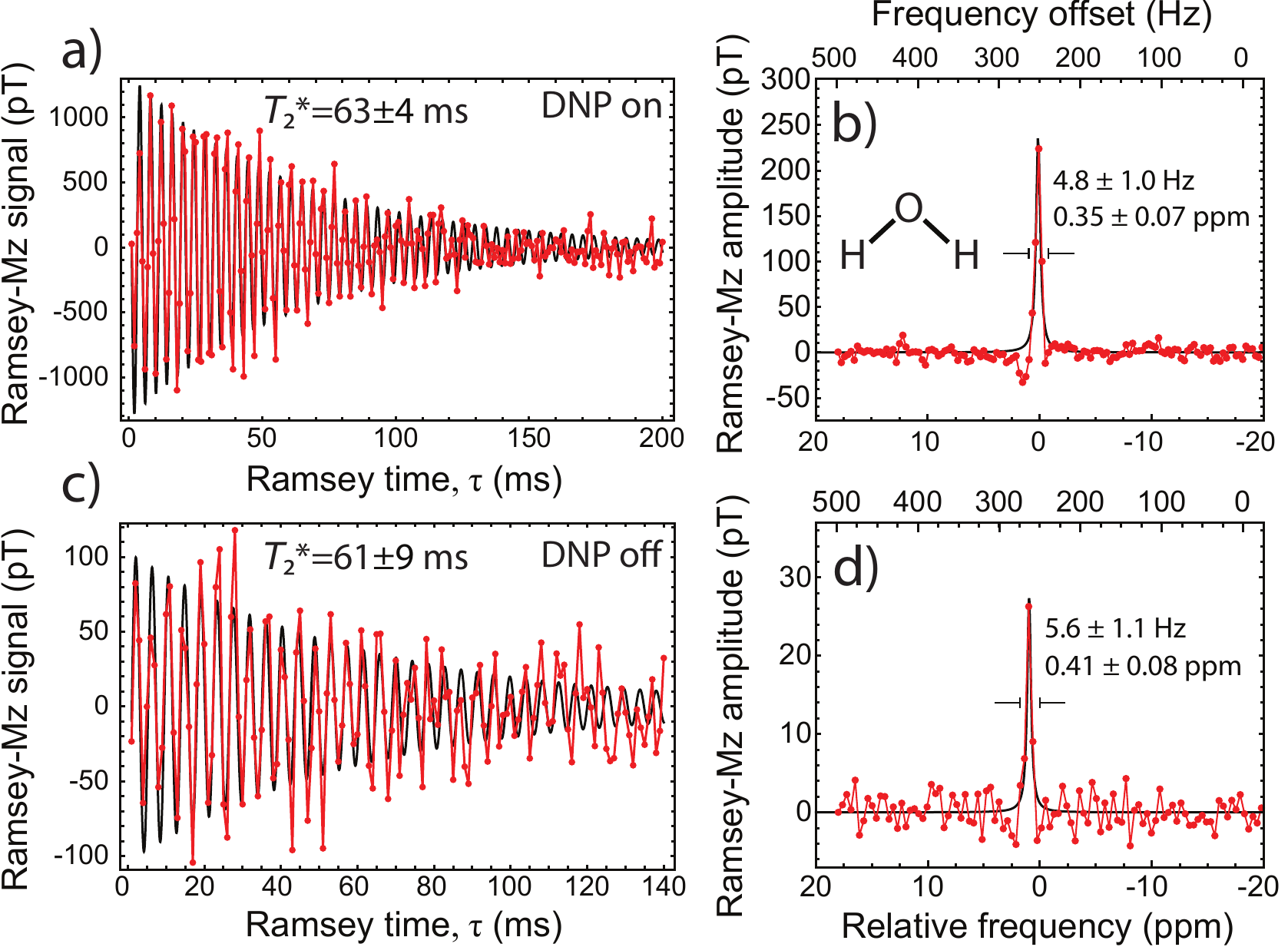}
    \caption{\textbf{Ramsey-$\bm{M_z}$ NMR spectroscopy of water.} (a) Time domain Ramsey-$M_z$ signal, $\overline{B_{\rm nuc}}(\tau)$, of water with Overhauser DNP turned on. A fit to an exponentially decaying sine function reveals the nuclear-spin dephasing time $T_2^{\ast}=63\pm4~{\rm ms}$. (b) NMR spectrum (\ref{sec:SI_RF_freqs}) obtained by taking the imaginary part of the Fourier transform of data in (a). A Lorentzian fit is used to extract the FWHM linewidth, $4.8\pm1.0~{\rm Hz}$, which corresponds to a fractional linewidth of $0.35\pm0.07~{\rm ppm}$. (c) Water Ramsey-$M_z$ signal without Overhauser DNP. The fitted nuclear-spin dephasing time is $T_2^{\ast}=61\pm9~{\rm ms}$. (d) NMR spectrum (\ref{sec:SI_RF_freqs}) obtained from the data in (c). The fitted FWHM linewidth is $5.6\pm1.1~{\rm Hz}$ ($0.41\pm0.08~{\rm ppm}$).}
    \label{fig:fig2_water}
\end{figure*}

\section{Results}

We first describe experiments with an analyte consisting of water doped with $2~{\rm mM}$ of TEMPOL electron-spin radicals. The TEMPOL spins serve two purposes: i) they reduce the proton $T_1$ to ${\sim}0.6~{\rm s}$, allowing for faster sequence repetition, and ii) they provide a means to hyperpolarize protons in water, via continuous-wave Overhauser DNP, thereby enhancing $M_z$.

Figure~\ref{fig:fig1p5} shows the diamond magnetometer signal, $B_{\rm nuc}(t)$, produced by hyperpolarized protons in water for the $\tau = 0$ case. RF $\pi$ pulses are applied every $1$ ms to modulate the sign of $M_z$. The envelope of the resulting magnetometer signal is fit with a Gaussian decay model $\vert B_{\rm nuc}(t) \vert = B_{\rm nuc,0}\, e^{-(t/T_{\rm meas})^2}$, where $T_{\rm meas}$ is a characteristic decay time due to a combination of nuclear spin relaxation and RF $\pi$ pulse imperfections. The ratio $T_{\rm meas}/(T_1+T_{\rm meas})$ sets an upper bound on the measurement duty cycle, $\delta$, for the protocol.

For a given value of $\tau$, a $B_{\rm nuc}(t)$ time trace, such as in Fig.~\ref{fig:fig1p5}, is processed by computing the average amplitude over a defined interval (${\sim}T_{\rm meas}$), denoted $\overline{B_{\rm nuc}}(\tau)$ (\ref{SI:sig_proc}). The process is repeated for different values of $\tau$ to build up a Ramsey interferogram.

Figure~\ref{fig:fig2_water}(a) shows the $\overline{B_{\rm nuc}}(\tau)$ Ramsey-$M_z$ signal for hyperpolarized protons in water, with detuning $\Delta_{\rm nuc}\approx-250~{\rm Hz}$. The data are fit to an exponentially-decaying sine function, yielding a decay time $T_2^{\ast}=63\pm4~{\rm ms}$. Figure~\ref{fig:fig2_water}(b) shows the imaginary part of the Fourier Transform of these data (\ref{SI:sig_proc}). A Lorentzian fit reveals a full-width-at-half-maximum (FWHM) linewidth of $4.8\pm1.0~{\rm Hz}$, which corresponds to a fractional spectral resolution of $350\pm70~{\rm ppb}$.  This is broader than that expected for water doped with TEMPOL at this concentration \cite{LEE2019, MAL2021}, likely due to uncompensated electromagnet field drifts. Nevertheless, it represents a ${\sim}$3-fold improvement in fractional spectral resolution compared to previous NV NMR experiments~\cite{SMI2019,ALL2022}.

The Overhauser DNP process introduces unwanted sample heating, and it is not broadly applicable to many analytes \cite{MAL2021,RAV2016}. Thus, for the remainder of this work we did not use it. Figure~\ref{fig:fig2_water}(c,d) shows the $\overline{B_{\rm nuc}}(\tau)$ Ramsey-$M_z$ signal and corresponding Fourier Transform of thermally polarized water. The spectral resolution is similar to that obtained with hyperpolarized water. By finite-element magnetostatic modeling (\ref{sec:SI_nucmag}), we estimate the maximum magnetic field amplitude is $\sim 150~{\rm pT}$ for room-temperature water protons in our experimental configuration. The observed signal amplitude in Fig.~\ref{fig:fig2_water}(c), $\overline{B_{\rm nuc}}(\tau=0)\approx100~{\rm pT}$, is slightly lower, likely due to RF $\pi$ pulse imperfections. 

Next we performed Ramsey-$M_z$ NV NMR spectroscopy on a sample of ethanol doped with $2~{\rm mM}$ of TEMPOL. Ethanol was chosen for its relatively simple chemical shift spectrum and its chemical compatibility with the adhesives used in our assembly. Moreover, ethanol was recently studied in several NV NMR proposals \cite{MUN2023,MUN2024,DAL2024}, facilitating a more direct comparison. The TEMPOL additive reduced the ethanol proton $T_1$ to ${\sim}0.6~{\rm s}$, improving the measurement duty cycle.

\begin{figure}[hbt]
    \centering
    \includegraphics[width=\columnwidth]{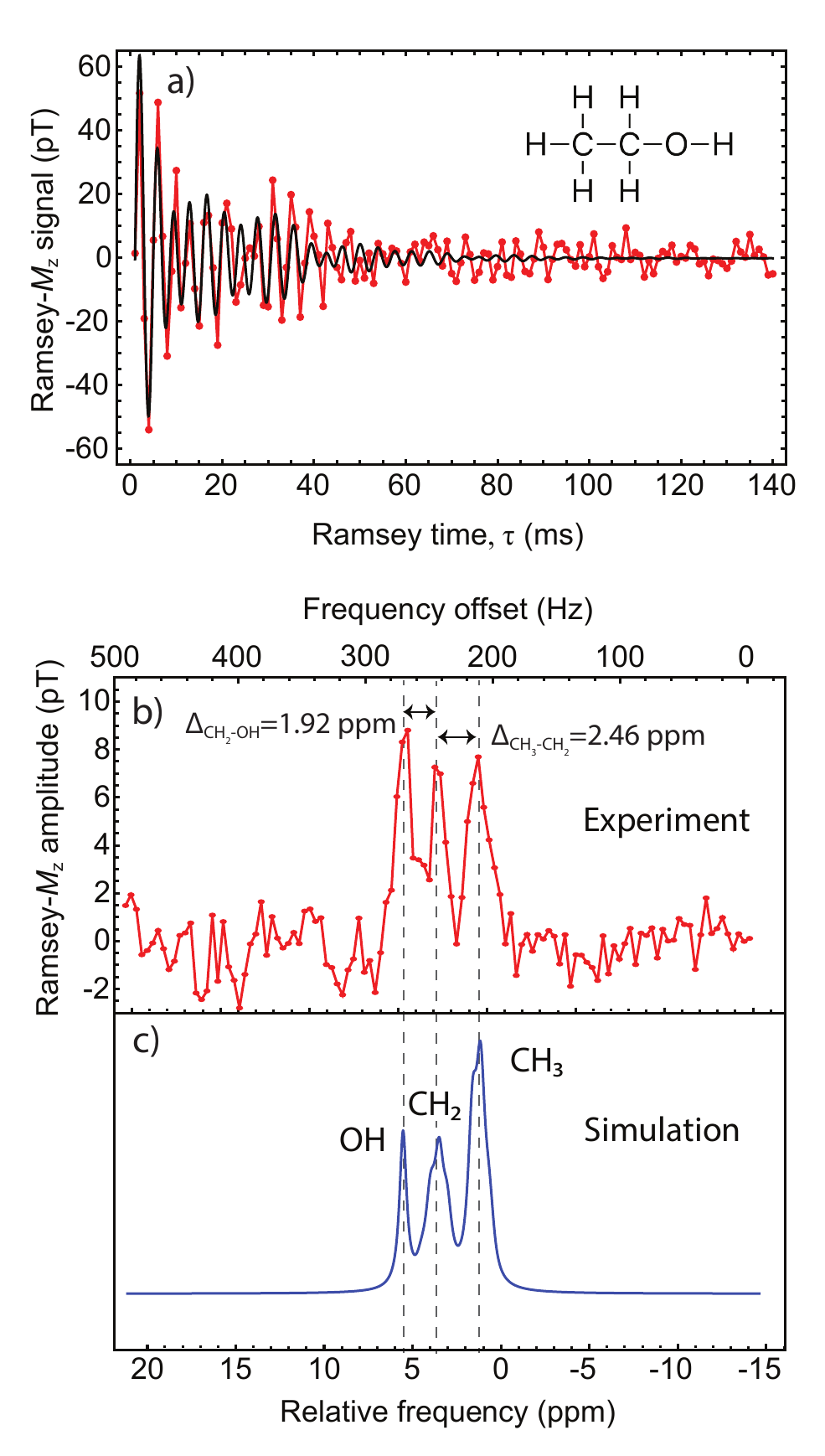}
    \caption{\textbf{$^{1}$H NMR spectroscopy of ethanol.} (a) Time-domain Ramsey-$M_z$ signal for ethanol, along with a fit to a three oscillator model. The data represent the average over ${\sim}1700$ Ramsey sweeps. (b) NMR spectrum of ethanol (\ref{sec:SI_RF_freqs}) obtained by taking the imaginary part of the Fourier transform of data in (a). Simulated~\cite{HOG2011} ethanol spectrum using known J-coupling and chemical shift parameters~\cite{SDB2024}. The spacings between peaks, $\Delta_{\rm CH_3\mbox{-}CH_2}=2.46~{\rm ppm}$, $\Delta_{\rm CH_2\mbox{-}OH}=1.92~{\rm ppm}$, align well with the experimental spectrum.}
    \label{fig:fig3_ethanol}
\end{figure}

Figure~\ref{fig:fig3_ethanol}(a) shows the $\overline{B_{\rm nuc}}(\tau)$ Ramsey-$M_z$ signal for protons in ethanol. Beats are observed, corresponding to interference of the methyl, ethyl, and hydroxyl $M_z$ signals. Figure~\ref{fig:fig3_ethanol}(b) shows the Fourier transform NMR spectrum. Three distinct peaks are resolved. The spacings between peaks, $\Delta_{\rm CH_3\mbox{-}CH_2}=2.46~{\rm ppm}$ and $\Delta_{\rm CH_2\mbox{-}OH}=1.92~{\rm ppm}$, correspond to the expected chemical shifts of the CH$_3$, CH$_2$ and OH chemical groups. 

To validate the results, we simulated the ethanol NMR spectrum under conventional $M_{\rm perp}$ inductive detection~\cite{HOG2011}, using the known J-coupling and chemical shift parameters for pure ethanol~\cite{SDB2024}, Fig.~\ref{fig:fig3_ethanol}(c). The relative peak positions in the experimental data match those predicted by the simulation. The integrated peak amplitudes are also in tolerable agreement, though the experimental values do not quite match the expected 3:2:1 ratio for CH$_3$:CH$_2$:OH, perhaps due to RF pulse errors. 

\section{\label{sec:Discussion}Discussion and outlook}
We have introduced a Ramsey-$M_z$ protocol that can perform high-resolution NV NMR spectroscopy at arbitrarily high magnetic field. This longitudinal-magnetization detection protocol has several fundamental advantages over those based on transverse magnetization, including a sensitivity that is largely independent of NMR frequency and a higher measurement duty cycle when $T_2^{\ast}\ll T_1$. We demonstrated the proof of concept at a magnetic field ($B_0=0.32~{\rm T})$ that is ${\sim}2$ times higher than previous high-resolution NV NMR experiments~\cite{BRU2023} and realized a fractional spectral resolution (${\sim}350~{\rm ppb}$ FWHM) that is ${\sim}3$ times better than previous works~\cite{SMI2019,ALL2022}. This allowed us to expand the range of analytes resolvable by NV NMR to ethanol, where we resolved the ${\sim}2~{\rm ppm}$ chemical-shift splittings. At $B_0=0.32~{\rm T}$, the fractional spectral resolution can, in principle, reach $\lesssim10~{\rm ppb}$, which is sufficient to resolve chemical shifts at a level rivaling conventional NMR spectrometers. The spectral resolution realized here was worse, but this was due to our use of an electromagnet to supply $B_0$, which can be replaced by the more stable field produced by permanent or superconducting magnets.

However, several challenges remain. First, in order to resolve J-coupling structure and chemical shifts in complex mixtures, operation at higher magnetic field, $B_0\gtrsim1~{\rm T}$, is necessary. Recently-proposed alternative NV NMR $M_z$ detection protocols~\cite{DAL2024,MUN2023} reported challenges due to spectral distortions when simulating NMR spectra at these fields. To investigate this for the Ramsey-$M_z$ protocol, we simulated the proton NMR spectrum of ethanol at $B_0=1~{\rm T}$ under typical pulse errors due to RF field inhomogeneity (\ref{sec:SI_trajectories}). 

\begin{figure}[htb]
    \centering
    \includegraphics[width=\columnwidth]{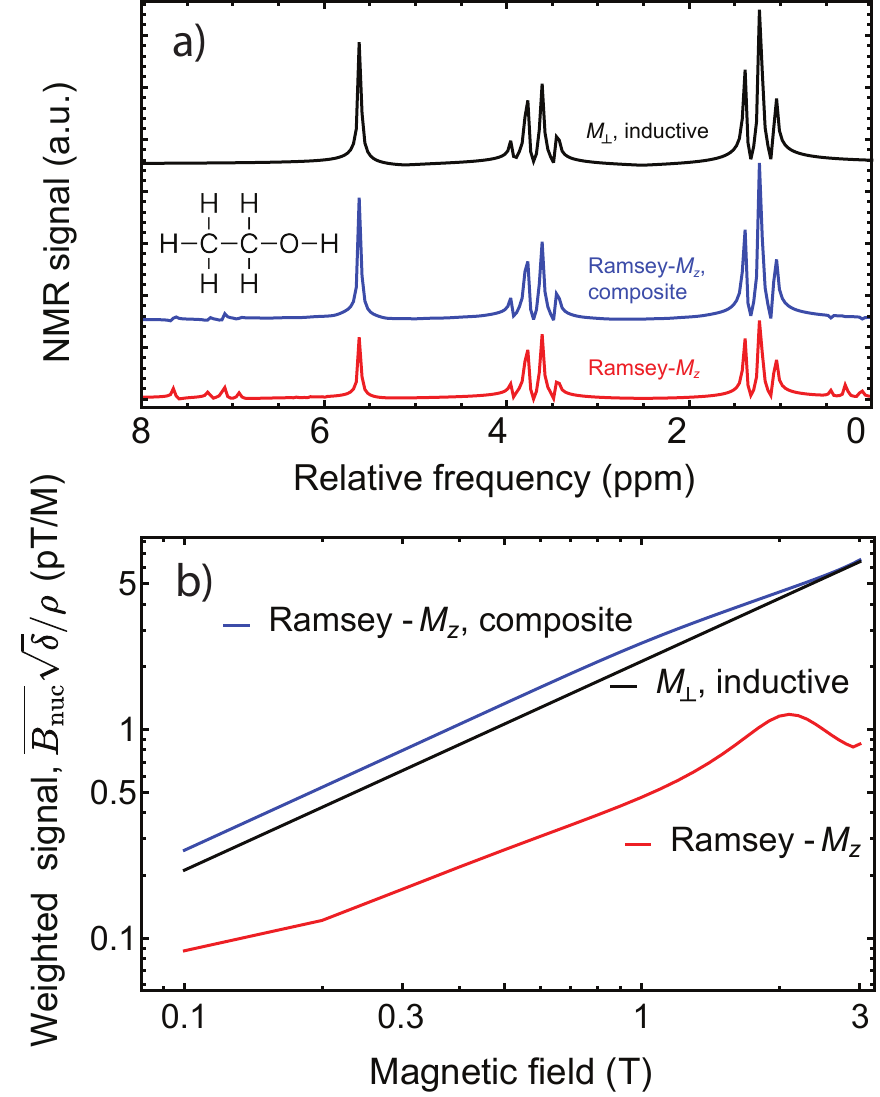}
    \caption{\textbf{High-field performance of Ramsey-$\bm{M_z}$ NMR.} (a) Simulated ethanol spectra at $B_0=1~{\rm T}$ for three detection protocols: $M_{\perp}$ detection with an inductive solenoid (black), Ramsey-$M_z$ NV NMR with the hard RF $\pi$ pulses used here (red), and Ramsey-$M_z$ NV NMR with phase-cycled composite Levitt-Freeman inversion pulses (blue). In each case, we assume the RF $B_1$ amplitude over the analyte volume has a Gaussian distribution with a mean that is $3\%$ lower than optimal and a standard deviation that is 5$\%$ of the mean (\ref{sec:SI_trajectories}). (b) Weighted NMR signal strength, $\overline{B_{\rm nuc}}\sqrt{\delta}/\rho$, of a cylindrical water sample (height=diameter=$h$) for $M_{\perp}$ inductive detection (black), Ramsey-$M_z$ NV NMR with composite inversion pulses (blue) and Ramsey-$M_z$ NV NMR with hard pulses (red). The model assumes the same $B_1$ error distribution as in (a), a 6 ppm RF detuning error, $T_1=5~{\rm s}$, and $T_2^{\ast}=0.1~{\rm s}$. The inductive model assumes unity fill factor and a perfect $\pi/2$ pulse. The Ramsey-$M_z$ model assumes a cylindrical NV sensor (height=$h/3$, diameter=$h$) underneath the analyte, with the NV axes and $\vec{B_0}$ normal to the sample-sensor interface. Details and a discussion of the shape of the Ramsey-$M_z$ curves are included in \ref{sec:SI_trajectories}.
    }
    \label{fig:fig4_future}
\end{figure}

Figure~\ref{fig:fig4_future}(a) shows simulated spectra for three cases: i) an ideal solenoid measurement of the transverse magnetization ($M_{\perp}$), ii) the Ramsey-$M_z$ protocol using the hard (rectangular) $\pi$ pulses employed here, and iii) a Ramsey-$M_z$ protocol where composite Levitt-Freeman inversion pulses~\cite{LEV1979,TYC1983} are used (\ref{sec:SI_trajectories}). Spectra simulated with the Ramsey-$M_z$ protocol are in general a close match to the $M_{\perp}$ inductive standard spectrum. This is in contrast to the alternative detection protocols explored in Refs.~\cite{DAL2024, MUN2023}, where truncations of either chemical shifts or scalar couplings are expected. In the case of Ramsey-$M_z$ with hard pulses, there are some spurious features and lineshape alterations that become more prominent as the RF pulse errors grow (\ref{sec:SI_trajectories}). However, the use of composite Levitt-Freeman inversion pulses restores agreement with the standard $M_{\perp}$ inductive spectrum, even in the presence of typical RF amplitude and detuning errors at $B_0\lesssim3~{\rm T}$.

Another important area for improvement is in sensitivity. The present sensor has a sensitivity of ${\sim}100~{\rm pT\,s^{1/2}}$, which is comparable to previous $M_{\perp}$ NV NMR works~\cite{GLE2018,SMI2019,BRU2023}, despite using a larger illumination volume. This sensitivity is inadequate for detection of small molecules at physiologically-relevant mM concentrations. Future experiments could use a larger $B_0$ to increase the net thermal polarization. We simulated the effective magnetic field produced by a proton solution as a function of $B_0$ using a combination of spin dynamics calculations and magnetostatic modeling (\ref{sec:SI_trajectories}). We assume the ideal geometry of $\vec{B_0}$, nuclear spin polarization, and NV center axes aligned normal to the sample-sensor plane, which doubles the signal compared to the in-plane geometry used here (\ref{SI:diamond_cut}). The analyte is modeled as a cylinder of equal diameter and height, $h$, with its axis along $\vec{B_0}$. The diamond sensor, placed beneath the analyte, is taken to be a cylinder of diameter $h$ and height $h/3$. We weight the analyte nuclear field by proton density ($\rho$) and measurement duty cycle, $\overline{B_{\rm nuc}}\sqrt{\delta}/\rho$, to allow for direct comparison to the sensor sensitivity. 

The simulated weighted NMR signal strength as a function of $B_0$ is shown in Fig.~\ref{fig:fig4_future}(b). For $M_{\perp}$ inductive detection, we assume a solenoid with unity fill factor, which results in a larger average $\overline{B_{\rm nuc}}$ than for the one-sided NV detection schemes. For hard pulses, the Ramsey-$M_z$ NV NMR protocol has a lower weighted signal strength in part because the duty cycle is reduced due to the interplay of imperfect inversion pulses and unwanted spin precession (\ref{sec:SI_trajectories}). However, for Levitt-Freeman composite inversion pulses, the duty cycle can exceed that of $M_{\perp}$ inductive detection, leading to a slightly higher weighted signal strength for $B_0\lesssim3~{\rm T}$.

The expected weighted signal strength for $B_0=3~{\rm T}$ under the Ramsey-$M_z$ protocol with composite pulses is ${\sim}7~{\rm pT/M}$. In \ref{SI:Ramsey_ENDOR}, we describe an improved detection protocol, ``Ramsey-ENDOR'', which embeds the RF inversion pulses inside NV Hahn echo sequences. This allows for use of high-sensitivity AC magnetometry schemes~\cite{TAY2008}. We predict a magnetometer sensitivity of ${\sim}0.5~{\rm pT\,s^{1/2}}$ is possible with a [111]-cut, $30{\mbox{-}}{\rm \upmu m}$-thick diamond illuminated by a $100{\mbox{-}}{\rm \upmu m}$-diameter laser beam. Using a repetitive readout protocol based on $^{14}$NV nuclear spin memory~\cite{JIA2009,NEU2010,PAG2014,SOS2021,ARU2023}, we predict a further improvement to ${\sim}0.1~{\rm pT\,s^{1/2}}$ is possible with the same sensor design. For a $0.7~{\rm nL}$ cylindrical analyte volume, the projected sensor performance corresponds to an NMR proton concentration sensitivity (SNR$=3$) of ${\sim}200~{\rm mM\,s^{1/2}}$ without repetitive readout and ${\sim}40~{\rm mM\,s^{1/2}}$ with it.

These concentration sensitivity estimates represent more than an order of magnitude improvement over conventional inductive detection of sub-nL volumes~\cite{STO1997,MCD2007,FUG2017,GRI2017}. While our protocol requires multiple nuclear polarization cycles to build up a spectrum, this allows for quantitative NMR spectroscopy with minimal distortion at high field~\cite{MUN2023,DAL2024}, and the number of cycles can be tailored to the analyte spectral complexity. Moreover, as a single-sided sensor, the NV NMR platform promises greater sample compatibility and lower susceptibility-related magnetic gradients (\ref{app:gradients}). By separating the nuclear precession and NV detection phases in time, our protocol avoids magnetic gradient broadening due to polarized NV spins, in contrast to transverse detection methods~\cite{GLE2018}. These attributes may facilitate wider adoption in chemical analysis and pharmaceutical science. 

In summary, we introduced a ``Ramsey-$M_z$'' detection protocol for high-resolution NV NMR spectroscopy that works at high magnetic fields with minimal spectral distortion. We demonstrated this technique by faithfully recording the chemical shift spectrum of ethanol. With improvements to the sensor design, we estimate a spectral resolution of ${\sim}1~{\rm ppb}$ and
concentration sensitivity of ${\sim}40~{\rm mM\,s^{1/2}}$ is feasible for sub-nanoliter analyte volumes.

\begin{acknowledgments}
We gratefully acknowledge advice and support from C.~Ramanathan, M.~Conradi, C.~Degen, P.~Kehayias, N.~Mosavian, A.~Berzins, and A.~Gravagne. \\
\textbf{Competing interests.} A.~J. is a co-founder of ODMR Technologies and has financial interests in the company. A.~F.~M. is the founder of NuevoMR LLC and has financial interests in the company. A.~J. and V.~M.~A. are co-inventors on related patents (US10914800B2, US11313817B2). The authors declare that they have no other competing interests.\\
\textbf{Author contributions.} A.~J., J.~S., and V.~M.~A. conceived the idea. J.~S., A.~J., and V.~M.~A. designed the experiments. J.~T.~D., Z.~P., Y.~S., and J.~S. built the main experimental apparatus. Y.~S., M.~D.~A, B.~A.~R., and M.~S.~Z. assisted with sample and sensor preparation and data collection. A.~F.~M. provided passive and active shimming of the electromagnet and advised on experimental design. J.~S. wrote control and automation software, acquired and analyzed the primary data, and wrote the initial manuscript draft. V.~M.~A. supervised the project. All authors helped edit the manuscript. \\
\textbf{Funding.} This work was supported by the National Science Foundation (CHE-1945148), National Institutes of Health (R41GM145129, DP2GM140921), and Moore Foundation (EPI-12968). 
\end{acknowledgments}

\clearpage
\appendix
\setcounter{equation}{0}
\setcounter{section}{0}
\makeatletter
\renewcommand{\thetable}{A\arabic{table}}
\renewcommand{\theequation}{A\Roman{section}-\arabic{equation}}
\renewcommand{\thefigure}{A\arabic{figure}}
\renewcommand{\thesection}{Appendix~\Roman{section}}
\makeatother

\section{Experimental layout}
\label{sec:SI_exp_layout}
\begin{figure}[htb]   
\includegraphics[width=0.65\columnwidth]{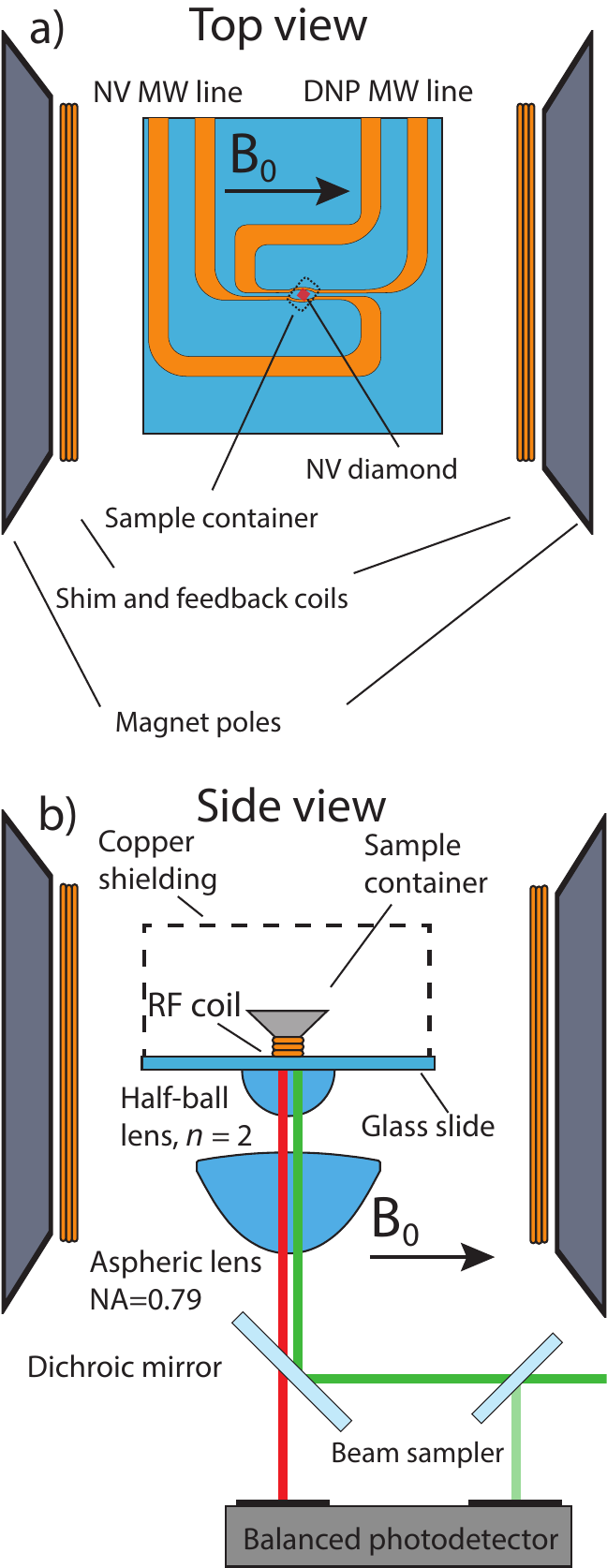}
    \caption{\textbf{Experimental setup.} (a) Top view of the experimental setup. The sample container is shown as a dashed line for clarity. (b) Side view of the setup. MW--microwave, NA--numerical aperture, $n$--refractive index.}
    \label{fig:SI_experimental_setup}
\end{figure}

Figure~\ref{fig:SI_experimental_setup} shows a detailed view of the experimental setup. To fabricate microwave lines, we first deposit a $10~{\rm nm}$ titanium adhesion layer and then a  $2~{\rm \upmu m}$ layer of copper onto a glass slide via a thermal evaporation process. Patterns are printed by ink toner and transferred to the copper-coated slides, forming a mask. The slides are then etched in ferric chloride and dipped in dilute hydrofluoric acid, leaving metal only under the printed traces.

The diamond membranes are adhered to the glass slide with a layer of Brewer Science ARC i-CON-16. This adhesive is chosen due to its high chemical resistance to the NMR fluid samples studied here. The photopolymer-printed analyte container is depicted in Fig.~\ref{fig:SI_experimental_setup}(b). At its base there is a ${\sim}1.5\times2~{\rm mm^2}$ rectangular constriction around which a small copper RF coil is wound. The sample container is adhered to the glass slide with the same ARC i-CON-16 compound as used for the diamond. At the bottom of the glass slide, we adhere a high index of refraction ($n=2$) $8~{\rm mm}$-diameter half-ball lens to improve the fluorescence collection efficiency. 

The assembly is enclosed in a copper tape housing (with openings left for optical access) to shield from stray RF interference. This is done to improve the SNR of the conventional NMR signal used for magnetic field shimming and other diagnostics. 

Fluorescence exiting the half-ball lens is collected by a Thorlabs ACL25416U (numerical aperture, NA=0.79) aspheric lens, passed through a dichroic mirror, and focused onto one channel of a Thorlabs PDB210A balanced photodetector. A beam sampler picks off some of the $532~{\rm nm}$ excitation laser beam, which is directed to the other channel of the photodetector to suppress common-mode laser power fluctuations.

\subsection{Microwave generation}
\label{sec:SI_MW_block}

\begin{figure}[b]
    \centering
    \includegraphics[width=\columnwidth]{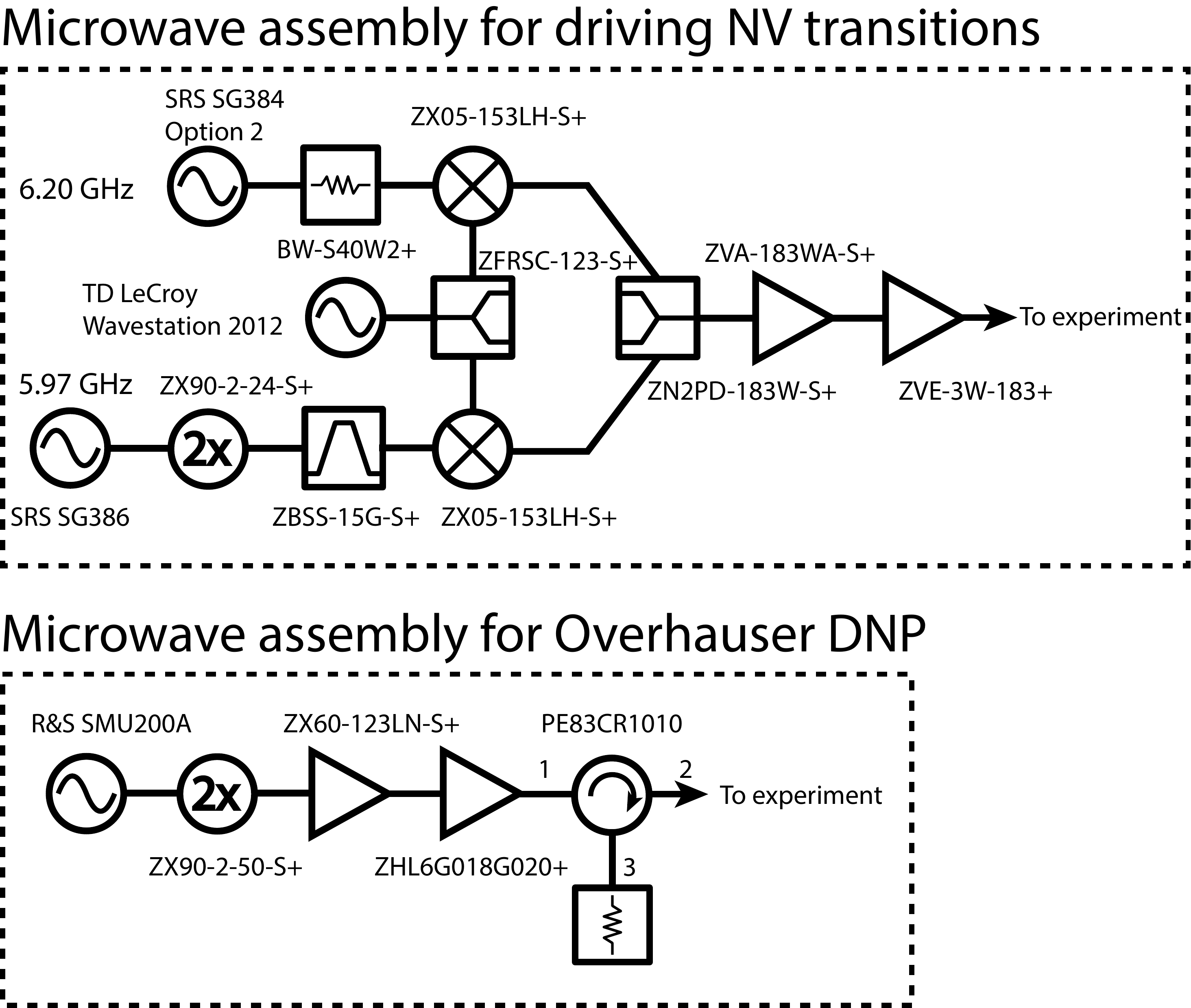}
    \caption{\textbf{Microwave generation block diagram.} (top) Block diagram for generating all the tones used to drive the NV transitions. (bottom) Block diagram for generating the Overhauser DNP MW drive. }
    \label{fig_sup:MW_synthesis}
\end{figure}

A block diagram of the microwave chain is shown in Fig.~\ref{fig_sup:MW_synthesis}. The diamond membrane is mounted in the middle of two separate copper traces--one for the (optional) Overhauser DNP drive and the other for driving NV spin transitions. Two SRS SG380 series microwave generators provide tones at each of the NV $f_{\pm}$ transition frequencies. The $f_-=6.20~{\rm GHz}$ tone is obtained from the frequency-doubled output of a SG384 (option 2) generator, while the $f_+=11.94~{\rm GHz}$ tone is reached by doubling the output signal of a SG386 generator and filtering out the fundamental frequency. A $40~{\rm dB}$ attenuator is placed in the $f_-$ path to compensate for losses on the $f_+$ path and match the microwave field amplitudes within the diamond.

To generate the six tones needed to drive each of the NV $f_{\pm,i}$ transitions, Eq.~\eqref{eq:fpm}, we mix a $A_{||}=2.17~{\rm MHz}$ signal, including a DC offset, with each of the $f_{\pm,0}$ tones. Thus, the $f_+$ path becomes three tones of equal amplitude at $f_{+,0}$ and $f_{+,0} \pm A_{||}$, and the $f_-$ path has equal-amplitude tones at $f_{-,0}$ and $f_{-,0}\pm A_{||}$. All tones are then combined using a broadband power combiner and amplified in two stages. 

In our experiments, $\gamma_{nv}B_0>D$, so both of the $f_{\pm}$ resonances shift in the same direction with respect to magnetic field, but in opposite directions with respect to temperature. Thus, when applying frequency modulation, we choose the phase of the frequency-modulation waveform to be the same for all six tones, which minimizes dependence on temperature while maximizing sensitivity to changes in $B_z$. This is in contrast to prior dual-resonance magnetometry experiments~\cite{WOJ2018,FES2020} which operated at low field ($\gamma_{nv}B_0<D$). There, the $f_{\pm}$ resonances shift in opposite directions with respect to magnetic field, but in the same direction with respect to temperature. To optimize those experiments for magnetometry, the frequency-modulation waveform for $f_+$ had a $\pi$ phase shift with respect to the frequency-modulation waveform for $f_-$.

To generate the Overhauser microwave drive at ${\sim}9.07~{\rm GHz}$, we used a Rohde \& Schwarz SMU200A signal generator and frequency doubled the output. The signal was amplified in two stages and passed through a circulator to avoid back-reflections into the amplifier.

\subsection{Magnetic field stabilization}
\label{SI:mag_stab}
The detection of the Ramsey-Mz NMR signal using CW-ODMR and a lock-in amplifier offers a simple and effective way to implement a magnetic field stabilization routine based on the same NV signal. The lock-in output voltage is directly proportional to shifts in $B_0$ experienced by the NV centers. Most $B_0$ drifts arise from the instability of the current source. The high inductance of the electromagnet acts as a low-pass filter, confining the majority of the magnetic field noise to frequencies that are much smaller than the modulated nuclear $M_z$ signal frequency (typically $1~{\rm kHz}$). With appropriate filtering of the lock-in feedback signal, it was possible to compensate for magnetic field drifts without suppressing the magnetic signal from the sample nuclei.

Stabilization was carried out by passing the lock-in voltage output through a Stanford Research Systems SR560 voltage preamplifier, configured to provide variable gain and a low-pass filter with a cutoff frequency of $30\,\mathrm{Hz}$ and a slope of $6\,\mathrm{dB/oct}$. The voltage preamplifier output was then routed to the amplitude modulation port of a Thorlabs LDC220C current controller, which powered a secondary pair of coils wrapped around the poles of the electromagnet. By selecting the appropriate polarity of the current in the secondary coils and optimizing the gain of the voltage preamplifier, we were able to stabilize $B_0$ to $\lesssim350~{\rm ppb}$ over the course of several hours.

\subsection{Experimental RF frequencies and determination of the analyte $T_1$}
\label{sec:SI_RF_freqs}
In water experiments, Fig.~\ref{fig:fig2_water} of the main text, the $\pi/2$ pulse frequency was $13.778000~{\rm MHz}$ while the $\pi$ pulse frequency was $13.778250~{\rm MHz}$. The observed NMR line position was $f_{\rm DNP} = 13.778251~{\rm MHz}$ for the DNP enhanced spectrum and $f_{\rm no DNP}=13.778263~{\rm MHz}$ for the water spectrum with no DNP.

For ethanol experiments, Fig.~\ref{fig:fig3_ethanol} of the main text, the $\pi/2$ pulse frequency was $13.778000~{\rm MHz}$ and the $\pi$ pulse frequency was $13.778240~\rm{MHz}$. The individual peak clusters were localized at $f_{\rm CH_3}=13.778210~{\rm MHz}$, $f_{\rm CH_2}=13.778243~{\rm MHz}$, and $f_{\rm OH}=13.778262~{\rm MHz}$.

The relaxation time of $T_1\approx 0.6~{\rm s}$ of the TEMPOL doped analytes was deduced by performing inductive detection and varying the repetition time between $\pi/2$ pulses. We observed that the coil NMR signal strength did not noticeably increase when the repetition time exceeded $4~{\rm s}$ and that the signal amplitude had dropped to $\sim 65\%$ when the repetition time was $0.6~{\rm s}$. Thus, in all Ramsey-$M_z$ experiments, the time between the final $\pi$ pulse of the modulated $M_z$ phase of the acquisition sequence and the first $\pi/2$ of the Ramsey phase of the acquisition sequence was fixed at $t_d=0.6~{\rm s}$.

In Fig.~\ref{fig:fig2_water}(b,d) of the main text, the ``frequency offset'' on the top axis is with respect to the $\pi/2$ pulse frequency (i.e. the $\pi/2$ pulse frequency is at $0~{\rm Hz}$). The ``relative frequency'' on the bottom axis is with respect to the $\pi$ pulse frequency (i.e. the $\pi$ pulse frequency is at $0~{\rm ppm}$). A small shift of $\Delta f = 12~{\rm Hz}$ is observed in the proton NMR peak location between the DNP on and DNP off spectra. This is due to a small change in the bias field between the two measurements. In Fig.~\ref{fig:fig3_ethanol}(b,c), the ``frequency offset'' axis is again with respect to the $\pi/2$ pulse frequency. The ``relative frequency'' axis is shifted such that the $\pi$ pulse frequency is equal to $3.69~{\rm ppm}$ to allow for comparison to the simulation (which sets $0~{\rm ppm}$ at the proton NMR frequency of tetramethylsilane.)

\section{Signal processing}
\label{SI:sig_proc}
\begin{figure}[ht]
    \centering
    \includegraphics[width=0.85\linewidth]{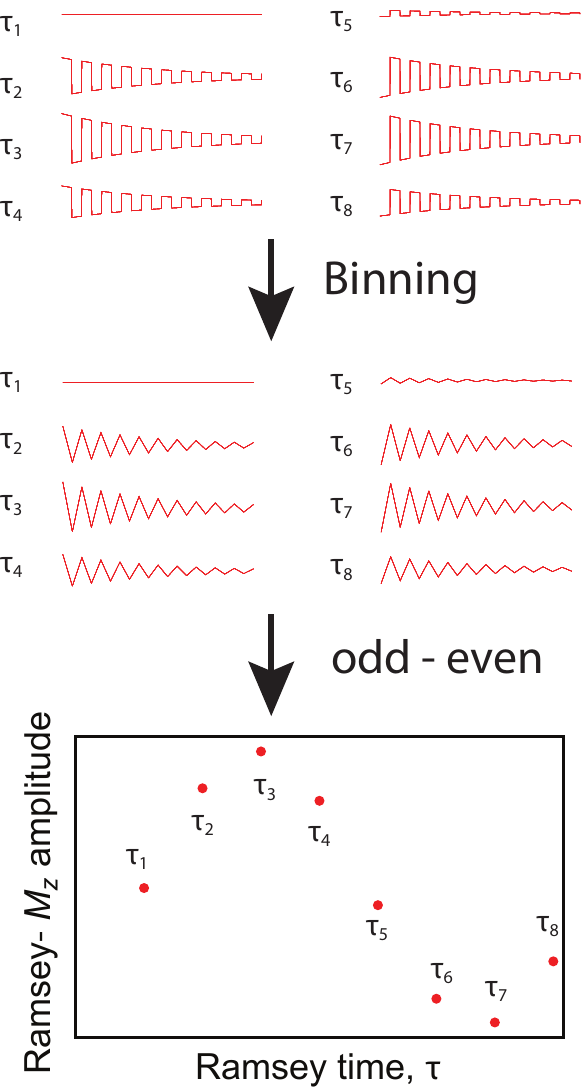}
    \caption{\textbf{Data processing schematic.} The raw data from the lockin amplifier consists of a series of waveforms measuring $B_{\rm nuc}(t)$ for each Ramsey delay $\tau$. The first step in processing the data is to average together data points over each interval between RF $\pi$ pulses. Each waveform is now oscillating up and down with a period of two points. The Ramsey-$M_z$ amplitude for a given value of $\tau$, $\overline{B_{\rm nuc}}(\tau)$, is computed by taking the mean of all samples at odd-indexed positions and subtracting the mean of all samples at even-indexed positions. 
    }
    \label{fig:dat proc}
\end{figure}

A raw Ramsey-$M_z$ dataset consists of an array of lock-in voltage time traces for each measured Ramsey $\tau$ value. Figure~\ref{fig:dat proc} shows a schematic of how we process these data. We first convert the voltage time traces to magnetic field units using the independent calibration process described in~\ref{SI:mag_calib}, resulting in a curve $B_{\rm nuc}(t)$ for each value of $\tau$. Next, we re-bin the magnetometer time traces by averaging together data points over each interval between RF $\pi$ pulses. Each $B_{\rm nuc}(t)$ trace now consists of a number of points equal to the number of $\pi$ pulses, oscillating up and down with a period of two points. 

The amplitude of a given $B_{\rm nuc}(t)$ signal is directly proportional to the magnetic field produced by the $M_z$ component of nuclear magnetization after the Ramsey sequence of delay $\tau$. To extract an average amplitude, $\overline{B_{\rm nuc}}$, we first apply a high-pass filter, with a cutoff frequency of ${\sim}400~{\rm Hz}$, to the re-sampled $B_{\rm nuc}(t)$ curves. Next, we multiply each $B_{\rm nuc}(t)$ curve by an exponential-decay windowing function:
\begin{equation}
\label{eq:window}
f(t)= \frac{T_{\rm acq}e^{-t/T_{\rm meas}}}{\int_0^{T_{{\rm acq}}} e^{-t'/T_{\rm meas}}dt'},
\end{equation}
where $T_{\rm meas}$ is the decay constant of the magnetization envelope (see Fig.~\ref{fig:fig1p5} of the main text) and $T_{\rm acq}$ is the length of each time trace. To produce the $\overline{B_{\rm nuc}}(\tau)$ curve in Fig.~\ref{fig:fig2_water}(a) in the main text, no windowing is used. For Fig.~\ref{fig:fig2_water}(c), we use $T_{\rm meas} = 55~{\rm ms}$. For Fig.~\ref{fig:fig3_ethanol}(a), we use $T_{\rm meas}= 35~{\rm ms}$. In all cases, $T_{\rm acq} = 100~{\rm ms}$. Finally, $\overline{B_{\rm nuc}}$ is computed by taking the mean of all samples at odd-indexed positions and subtracting the mean of all samples at even-indexed positions.

In order to produce NMR spectra, we add a small amount of optional zero-padding to the end of each $\overline{B_{\rm nuc}}(\tau)$ curve. In Fig.~\ref{fig:fig2_water}(b) of the main text, we zero-pad out to $\tau_{\rm max}=250~{\rm ms}$. For Fig.~\ref{fig:fig2_water}(d), we zero-pad out to $\tau_{\rm max}=190~{\rm ms}$. For Fig.~\ref{fig:fig3_ethanol}(b), we zero-pad out to $\tau_{\rm max}=240~{\rm ms}$. Next, we apply the exponential decay windowing function of Eq.~\eqref{eq:window}, except $\tau_{\rm max}$ replaces $T_{\rm acq}$ and an empirical time constant, $T_{\rm ram}$ replaces $T_{\rm meas}$. The values of $T_{\rm ram}$ are selected to improve the SNR of a given spectrum without significantly altering the lineshapes (i.e. the spectral lineshapes and FWHM appear similar to those without windowing). For Fig.~\ref{fig:fig2_water}(b), we do not apply this windowing step. For Figs.~\ref{fig:fig2_water}(d) and \ref{fig:fig3_ethanol}(b), we apply $T_{\rm ram}=75~{\rm ms}$. 

Finally, an NMR spectrum is then obtained by taking the Fourier transform of the padded and windowed $\overline{B_{\rm nuc}}(\tau)$ signal and choosing the ``appropriate'' phase. In our case, we select the purely imaginary component, which is a result of detuning the Ramsey drive field by ${\sim}250~{\rm Hz}$ and starting the acquisition at $\tau=1~{\rm ms}$. A different choice in detuning frequency and initial $\tau$ value would have changed the phase of the signal.

\section{Experimental geometry considerations for $M_z$ detection}
\label{SI:diamond_cut}
The experimental geometry--including the spatial arrangement of the diamond and analyte, the crystallographic cut of the diamond, and the direction of the magnetic field--has a strong impact on the expected nuclear field, $B_{\rm nuc}$, detected by the diamond magnetometer. While $B_{\rm nuc}$ can be computed numerically using magnetostatic calculations for any geometry, here we provide an analytical treatment of a simplified geometry to provide intuition for the optimal choices. 

We assume the analyte is a cylindrical volume, $V$, directly on top of a cylindrical NV illumination volume, with both cylinders sharing the same symmetry axis. We take the limit that the NV illumination cylinder is much smaller than the analyte volume, such that it can be effectively be treated as a point detector. Moreover, we assume the analyte's longitudinal magnetization, $\vec{M}$, is parallel to the NV axis (described as a unit vector $\hat{\epsilon}_{nv}$), which is equivalent to the usual configuration of $\vec{B_0}$ being applied along the interrogated NV axis.

The nuclear magnetic field projection on the NV axis can be expressed as:
\begin{equation}
\label{eq:magfield_nv}
    \vec{B}_{\rm nuc}\cdot\hat{\epsilon}_{nv} = \hat{\epsilon}_{nv} \cdot \frac{\mu_0}{4\pi} \int\limits_V \left(\frac{3 \vec{r}\,(\vec{M} \cdot \vec{r})}{r^5}-\frac{\vec{M}}{r^3}\right) dV ,
\end{equation}
where $\mu_0$ is the vacuum permeability, $\vec{M} = M_0\, \hat{\epsilon}_{nv}$ is the nuclear magnetization vector, and $\vec{r}$ is the displacement vector (magnitude $r$) between the NV sensor and a volume element of the analyte. 

We consider three possible choices of $\hat{\epsilon}_{nv}$. The first, denoted ``(100)'', corresponds to a (100)-cut diamond where the interrogated NV axis lies in the $xz$ plane. The second, ``(110)'', corresponds to a (110)-cut diamond where the interrogated NV axis is along the $x$ axis, in the plane of the diamond faces.  The third, ``(111)'', corresponds to a (111)-cut diamond where the interrogated NV axis lies along the $z$ axis, normal to the diamond faces. Expanding Eq.~\eqref{eq:magfield_nv} in Cartesian coordinates for each case, we find:
\begin{equation}
\begin{aligned}
    \vec{B}_{\rm nuc}\cdot\hat{\epsilon}_{nv} & = \\
    {\rm (100)} &:~\frac{M_0 \upmu_0}{4\pi}\int\limits_V \frac{x^2-y^2+2\sqrt{2} x z}{\left(x^2+y^2+z^2\right)^{5/2}} dxdydz\\
    (110)  & :~\frac{M_0 \upmu_0}{4\pi}\int\limits_V \frac{2x^2-y^2-z^2}{\left(x^2+y^2+z^2\right)^{5/2}} dxdydz \\
        (111)  &:~\frac{M_0 \upmu_0}{4\pi}\int\limits_V \frac{2 z^2 - x^2 - y^2}{\left(x^2+y^2+z^2\right)^{5/2}} dxdydz.
\label{eq:cartesian}
\end{aligned}
\end{equation}
Rewriting Eq.~\eqref{eq:cartesian} in a cylindrical coordinate system, we obtain:
\begin{equation}
\begin{aligned}
    \vec{B}_{\rm nuc}\cdot\hat{\epsilon}_{nv} & = \\
        {\rm (100)} &:~\frac{M_0 \upmu_0}{4\pi}\int\limits_V \frac{2\sqrt{2}r^2 z\cos\left(\phi\right)+r^3\cos\left(2\phi\right)}{(r^2+z^2)^{5/2}} dr dz d\phi \\
    (110)&:~ \frac{M_0 \upmu_0}{4\pi}\int\limits_V \frac{r^3-2z^2r+3r^3\cos\left(2\phi\right)}{2(r^2+z^2)^{5/2}} dr dz d\phi \\
        (111)&:~\frac{M_0 \upmu_0}{4\pi}\int\limits_V \frac{2z^2r-r^3}{(r^2+z^2)^{5/2}} dr dz d\phi.
\label{eq:cylindrical}
\end{aligned}
\end{equation}
Finally, we can carry out the integration over $\phi$ from $0$ to $2\pi$ in Eq.~\eqref{eq:cylindrical}. Taking into account that the integrals of $\cos{\phi}$ and $\cos{2\phi}$ are both zero, we find:
\begin{equation}
\begin{aligned}
    \vec{B}_{\rm nuc}\cdot\hat{\epsilon}_{nv} & = \\
    {\rm (100)} &:~0\\
    (110)&:~ \frac{M_0 \upmu_0}{4\pi}\iint - \frac{\pi r\left(2 z^2-r^2\right)}{(r^2+z^2)^{5/2}} dr dz \\
        (111)&:~\frac{M_0 \upmu_0}{4\pi}\iint \frac{2\pi r \left(2z^2-r^2\right)}{(r^2+z^2)^{5/2}} dr dz.
\label{eq:cylindricalanswer}
\end{aligned}
\end{equation}

The two integrands in Eq.~\eqref{eq:cylindricalanswer} differ only by a sign and a constant factor of $2$, with the magnitude of $\vec{B}_{\rm nuc}$ being larger in the (111) case. While this result was obtained for the specific case of a cylindrically symmetric analyte, numerical simulations suggest that this trend holds for other sample geometries as well, as long as the NV illumination volume is not much larger than the analyte volume and the shape of the analyte volume is roughly isometric (i.e. none of the dimensions is much larger than any other).

\section{Magnetostatic calculations}

\subsection{Nuclear magnetic field, $B_{\rm nuc}$, estimate}
\label{sec:SI_nucmag}
To estimate the magnetic field within the diamond illumination region due to polarized protons in the analyte, we use finite-element magnetostatic modeling. The magnetization of thermally-polarized nuclear spins is:
\begin{equation}
    M = n\mu \,\text{tanh}\left(\frac{\mu B_0}{k_B T}\right),
    \label{eq:nucmag}
\end{equation}
where $n=6.7\times 10^{28}~{\rm m^{-3}}$ is the number density of protons in water, $\mu = 1.4 \times 10^{-26}~{\rm J/T}$ is the proton magnetic moment, $B_0= 0.32~{\rm T}$ is the bias magnetic field, $T\approx300~{\rm K}$ is the temperature, and $k_B=1.38\times10^{-23}~{\rm J/K}$ is the Boltzmann constant. Inserting these values into Eq.~\eqref{eq:nucmag}, we find the volume magnetization is $M_0\approx 1.0\times 10^{-3}~{\rm J/(T\,m^{3})}$.

\begin{figure}[hbt]
    \centering
    \includegraphics[width=0.8\columnwidth]{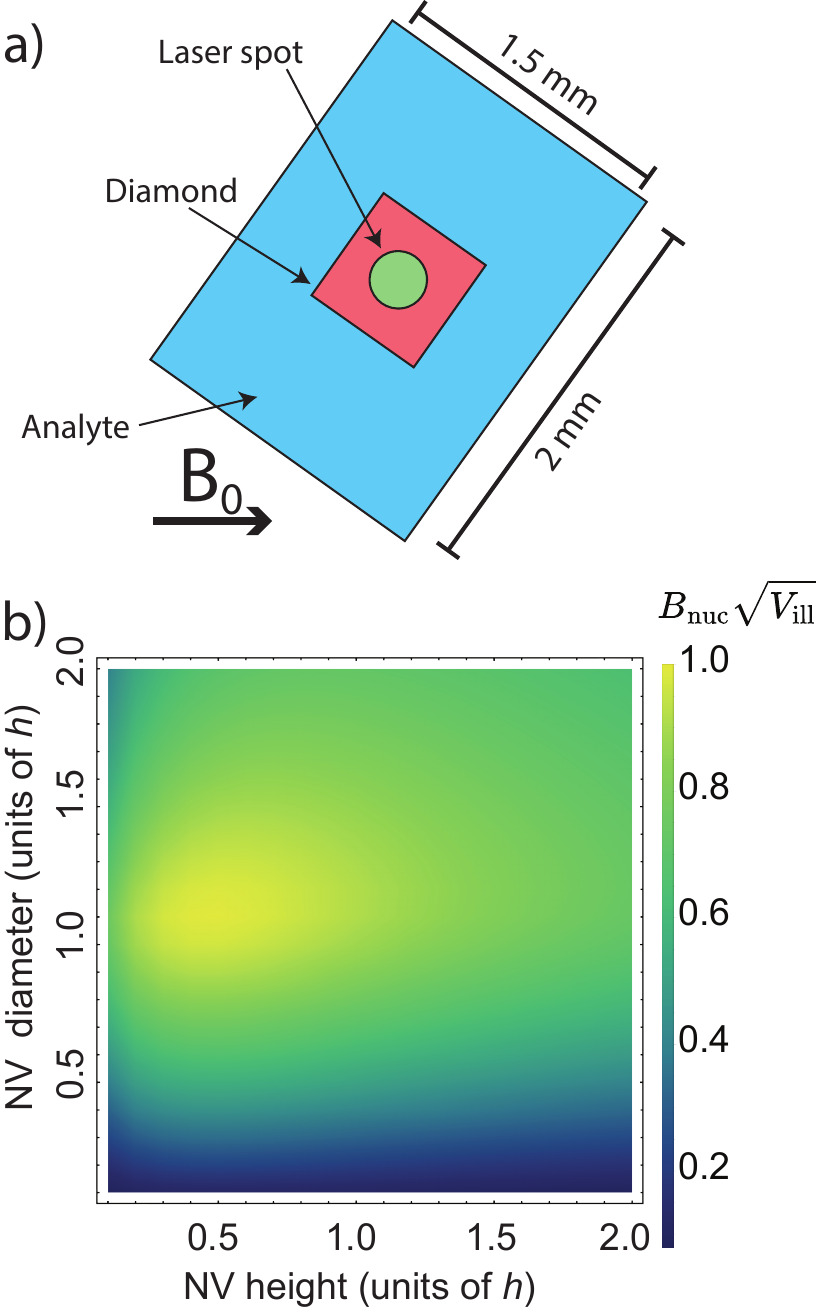}
    \caption{\textbf{Model geometry and optimal sensor size.} (a) Sensor-sample geometry. The diamond has dimensions dimensions ${\sim}250{\times}250{\times}60~{\rm \upmu m^3}$ with (110) faces. The $\vec{B_0}$ field is aligned along one of the in-plane NV axes, forming a $\sim 55\degree$ angle with one of the edges. The laser beam (diameter: ${\sim}135~{\rm \upmu m}$) is positioned roughly in the middle of the diamond. The $2\times1.5\times1.5~{\rm mm^3}$ sample holder is centered about the diamond and defines the shape of the analyte. The analyte makes direct contact with the diamond. (b) NMR SNR figure of merit, $B_{\rm nuc}\sqrt{V_{\rm ill}}$, as a function of the height and diameter of a cylindrical NV illumination region. The analyte is assumed to be a cylinder with ${\rm diameter}={\rm height}=h$ directly above the NV illumination region. }
    \label{fig_sup:SI_geom}
\end{figure}

 The volume of the analyte is defined by a 3D printed sample container and can be approximated by a block with dimensions $2\times1.5\times1.5~{\rm mm^3}$. The relative positions of the sample holder, diamond, and the illuminated portion of the diamond are depicted in Fig.~\ref{fig_sup:SI_geom}(a). With this geometry and estimate for $M_0$, we calculate an upper bound for the average nuclear magnetic field over the NV illumination volume of $B_{\rm nuc,0} \approx 270~{\rm pT}$. This corresponds to the maximum possible nuclear magnetic field for zero Ramsey delay, $\tau=0$.

In a Ramsey-$M_z$ experiment, the modulated nuclear field amplitude, $|B_{\rm nuc}(t)|$, decays with a Gaussian envelope on a timescale of $T_{\rm meas}$, see Fig.~\ref{fig:fig1p5} of the main text. To determine the Ramsey-$M_z$ signal in Fig.~\ref{fig:fig2_water}, for a given value of $\tau$, we compute the average amplitude over an interval of length $1.5T_{\rm meas}$. Applying the same process to our simulated upper-bound nuclear field, we find an upper-bound on the time-averaged amplitude (for $\tau=0$) of $\overline{B_{\rm nuc}}\approx150~{\rm pT}$.

\subsection{Estimate of the analyte sensing volume}
\label{sec:SI_sens_vol}
The magnetic field produced by a distant nuclear spin at the sensor falls off as $1/r^3$, where $r$ is the distance between the spin and the sensor. Here ``distant'' is taken to mean that $r$ is larger than any of the physical dimensions of the sensor. In cases where the total analyte volume is much larger than the sensor, this means that the nuclei closest to the sensor have an outsized contribution to the total observed signal strength compared to the bulk of the analyte. To account for this, an effective sensing volume, $V_{\rm sens}$, is defined as the volume of analyte which contributes $50\%$ of the signal strength that would be produced by the entire analyte volume~\cite{STA2013,GLE2018,SMI2019}. 

To estimate $V_{\rm sens}$, we take the entire analyte volume to be the $2\times1.5\times1.5~{\rm mm^3}$ volume around which the RF coil is wrapped.  
We performed finite element magnetostatic simulations to determine the nuclear field amplitude, $B_{\rm nuc,0}$, averaged over the NV illumination volume for a given analyte volume. We performed these simulations for different analyte volumes and compared to the value $B_{\rm nuc,0}\approx270~{\rm pT}$ obtained for the entire analyte volume. We find that a cylinder of radius $r=75~{\rm \upmu m}$ and height $h=55~{\rm \upmu m}$ directly above the illuminated spot on the diamond produces a signal $B_{\rm nuc,0}\approx135~{\rm pT}$. This analyte volume corresponds to $V_{\rm sens}=1~{\rm nL}$.  

\subsection{Optimal sensor dimensions}
\label{SI:opt_dim}
In Fig.~\ref{fig:fig4_future}(b) and related discussion section of the main text, we assume an ideal NV illumination region of diameter $h$ and height $h/3$ is used to detect an analyte cylinder of ${\rm diameter}={\rm height}=h$. These ideal NV illumination dimensions are selected based on optimization from magnetostatic modeling. To find them, we assumed the analyte is a cylinder of diameter=height$=h$, placed directly above a cylindrical diamond sensor, with the two cylinders' axes aligned. We further assumed $\vec{B_0}$ and the interrogated NV axis lie along the cylinder axis, corresponding to a (111)-cut diamond. The diamond magnetometer sensitivity scales as $1/\sqrt{V_{\rm ill}}$ (see~\ref{SI:Ramsey_ENDOR}), where $V_{\rm ill}$ is the NV illumination volume. Thus, we define an SNR figure of merit as $B_{\rm nuc}\sqrt{V_{\rm ill}}$, where $B_{\rm nuc}$ is the analyte nuclear field amplitude averaged over the NV illumination region. Figure~\ref{fig_sup:SI_geom}(b) shows a map of the SNR figure of merit as a function of NV illumination dimensions. The region of optimal dimensions is relatively broad, and the illumination dimensions of diameter=$h$ and height=$h/3$ are nearly optimal.

\subsection{NMR line broadening due to mismatched magnetic susceptibilities}
\label{app:gradients}

In the main text, we argued that the single-sided geometry of the NV NMR sensor allows for superior spectral resolution when compared to microcoil NMR. The reason is due to the potential for more gradual magnetic gradients, owing to the planar geometry. In the case of a microcoil~\cite{DAV2021,PAQ2020}, there are necessarily sharp discontinuities in magnetic susceptibility present between the coil and sample. In the single-sided diamond sensor, the discontinuities in susceptibility are more manageable.

\begin{figure}[h]
    \centering
    \includegraphics[width=0.86\linewidth]{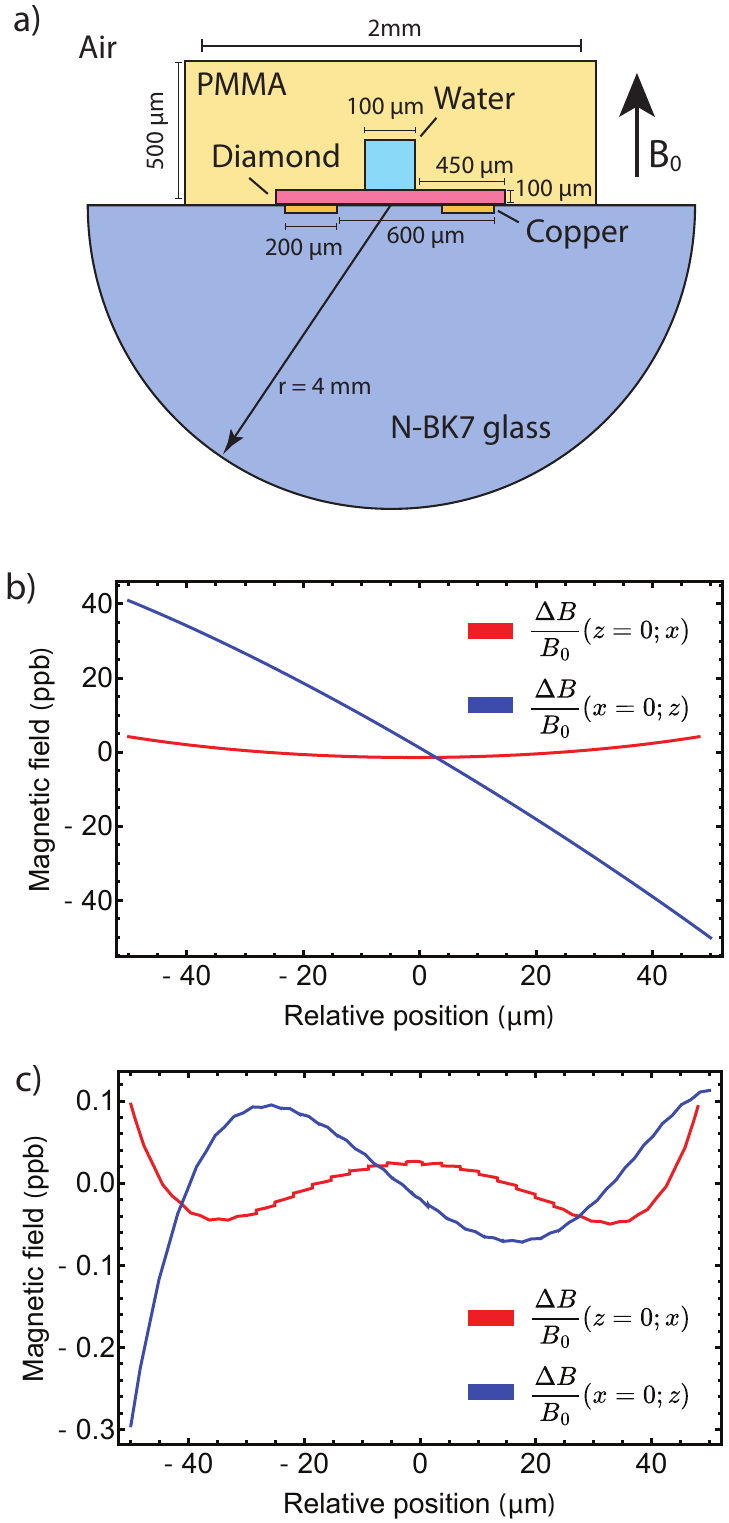}
    \caption{\textbf{Susceptibility-induced gradients.} (a) Device schematic used for gradient estimates. $\vec{B_0}$ is assumed to be normal to the diamond faces. Volume susceptibility of component materials are shown in Table~\ref{tab:SI_grads}. RF coils addressing analyte nuclei are assumed to be far from the analyte and are thus neglected. (b) Relative shift in magnetic field in the sample region along $x$ (parallel to the diamond) and $z$ (perpendicular to the diamond). (c) Remaining shift in magnetic field after removing 1st and 2nd order gradients. Small discontinuities in the curves are due to simulation mesh effects.}
    \label{fig:SI_grad_sup}
\end{figure}

To numerically model the field homogeneity, we assume an idealized microfluidic geometry shown in Fig.~\ref{fig:SI_grad_sup}(a). We use a two-dimensional simulation of the magnetostatics under an applied field, which is a reasonable approximation when the dimension of the microfluidic channel out of the page is much larger than its cross section. Figure~\ref{fig:SI_grad_sup}(b) shows the relative shift in magnetic field across the sample volume (i.e. the microfluidic channel's cross section) due to the different susceptibility of sensor materials (Table~\ref{tab:SI_grads}). The gradients exceed our desired $1~{\rm ppb}$ tolerance, but they are gradual and can thus be canceled out by standard magnetic field shims. 

\renewcommand{\arraystretch}{1.2}
\setlength\extrarowheight{0pt}
\begin{table}[htb]
    \centering
    \begin{tabular}{|c|c|}
    \hline
      Material & Volume susceptibility \\
        \hline
        Diamond  & $-2.2\times10^{-5}$ \\
        \hline 
        PMMA & $-9.06\times10^{-6}$\\
        \hline 
        Copper & $-9.6\times10^{-6}$ \\
        \hline 
        N-BK7 glass \cite{WAPLER2014} &  $-1.3\times 10^{-5}$\\
        \hline
        Air & $0$\\
        \hline 
        Water & $-9.04\times10^{-6}$\\
        \hline
        
    \end{tabular}
    \caption{\textbf{Magnetic susceptibility of device materials.}}
    \label{tab:SI_grads}
\end{table}
\renewcommand{\arraystretch}{1}

To mimic the effect of 1st- and 2nd-order shimming, the field profiles in Fig.~\ref{fig:SI_grad_sup}(b) were fit by a second order polynomial and the fit polynomial was subtracted from the field profile. Fig.~\ref{fig:SI_grad_sup}(c) shows the resulting profiles, which depict the magnetic gradients under ideal 1st- and 2nd-order shimming. It is seen that the field deviations are well below $1~{\rm ppb}$. Assuming a bias field value of $B_0 = 3~{\rm T}$, the required 1st- and 2nd-order gradient shim field strengths are $\sim 20~{\rm \upmu T/cm}$ and $\sim 1~{\rm mT/cm^2}$, respectively. These are moderate values for typical room-temperature shims, suggesting the approach is feasible.

\section{Diamond magnetometer sensitivity}

\subsection{Magnetic sensitivity calibration}
\label{SI:mag_calib}

To calibrate the sensitivity of the NV diamond magnetometer, we apply a test signal through a single loop wound of AWG 18 magnet wire. The current is applied through a Thorlabs LDC210C current supply which is amplitude modulated by a signal generator (Teledyne Lecroy Wavestation 2012). The current-to-field conversion factor is determined by applying a DC current to the coil and monitoring the position of the water NMR peak. In this case, the NMR detection is done inductively using the same RF coil that is used to drive analyte nuclear spins, see Fig.~\ref{fig:fig1}(a). Next, we use the calibrated testfield coil to apply a square wave magnetic field at $1~{\rm kHz}$, which is the same frequency as the modulated $M_z$ signals used in the Ramsey-$M_z$ protocol. We record the diamond magnetometer signal output from the lock-in amplifier and process the time trace in the same manner as described in \ref{SI:sig_proc}. This allows us to convert from the diamond magnetometer lock-in signal (mV) to NV sensor signal (pT). This conversion factor is applied when recording $B_{\rm nuc}(t)$ signals, as shown on the two vertical axes of Fig.~\ref{fig:fig1p5} of the main text. 
 
To obtain an estimate of the diamond magnetometer noise floor, we repeat this process several times, with the test signal off, and calculate the standard deviation of measurements over ${\sim}1~{\rm s}$ intervals. The standard deviation is converted to magnetic field sensitivity units using the independently-determined conversion factor. The typical sensitivity, ${\sim}100~{\rm pT \,s^{1/2}}$, indicates that a sinusoidal $1~{\rm kHz}$ magnetic field of amplitude $100~{\rm pT_{rms}}$ could be detected with SNR$\approx1$ after $1~{\rm s}$ of integration.

\subsection{Comparison to fundamental limits}
\label{sec:SI_sens}
 
As discussed in the main text and above, the sensitivity of the present diamond magnetometer is ${\sim}100~{\rm pT\,s^{1/2}}$. The observed sensitivity is mostly accounted for by contributions due to photoelectron shot noise and microwave phase noise.

In the photoelectron shot noise limit, the sensitivity is given by:
\begin{equation}
\label{eq:psn}
\eta_{\rm psn}\approx\frac{\sqrt{2}\,\Gamma}{\gamma_{nv} \,C\sqrt{I_0/q}},
\end{equation}
where the factor of $\sqrt{2}$ is due to our use of balanced detection, $\Gamma\approx0.8~{\rm MHz}$ is the ODMR FWHM linewidth, $C\approx0.03$ is the ODMR contrast, $I_0\approx70~{\rm \upmu A}$ is the mean photocurrent, and $q$ is the electron charge. There are additional factors due to our use of lock-in detection and details of the ODMR lineshape that are omitted in Eq.~\eqref{eq:psn}. Their combined contributions comprise a pre-factor of ${\sim}1$. Inserting the experimental values into Eq.~\eqref{eq:psn} results in the estimate $\eta_{\rm psn}\approx65~{\rm pT\,s^{1/2}}$.

Microwave phase noise is also expected to play a contributing role, particularly since we use relatively high NV transition frequencies~\cite{BER2024}. Two separate SRS SG380 generators were used to produce the microwave drive for the $f_{\pm}$ transitions. Using the estimates for a CW-ODMR magnetometer in Ref.~\cite{BER2024}, and assuming a doubling of phase noise with each frequency-doubling stage, we estimate the equivalent noise floor due to the impact of microwave phase noise to be $\eta_{\rm \phi n}\approx30~{\rm pT\, s^{1/2}}$.

\section{The effect of RF pulse errors}
\label{sec:SI_trajectories}

RF pulse imperfections can be an important source of systematic error in the Ramsey-$M_z$ protocol. These are typically due to a combination of: i) inhomogeneity in the RF field amplitude, $B_1$, which leads to deviations from ideal nuclear-spin inversion pulses, ii) the presence of chemical shifts, which lead to a variation in resonance frequency, resulting in some spins being driven off resonance. During the modulated $M_z$ phase of the Ramsey-$M_z$ protocol, Fig.~\ref{fig:fig1}(c), a large number of RF inversion pulses are applied, interleaved with periods of free evolution, so the effect of pulse imperfections tends to accumulate. Moreover, the RF inversion pulses have a ``refocusing'' effect, which prolongs the spin dynamics.

\begin{figure}[hbt]
    \centering
    \includegraphics[width=\linewidth]{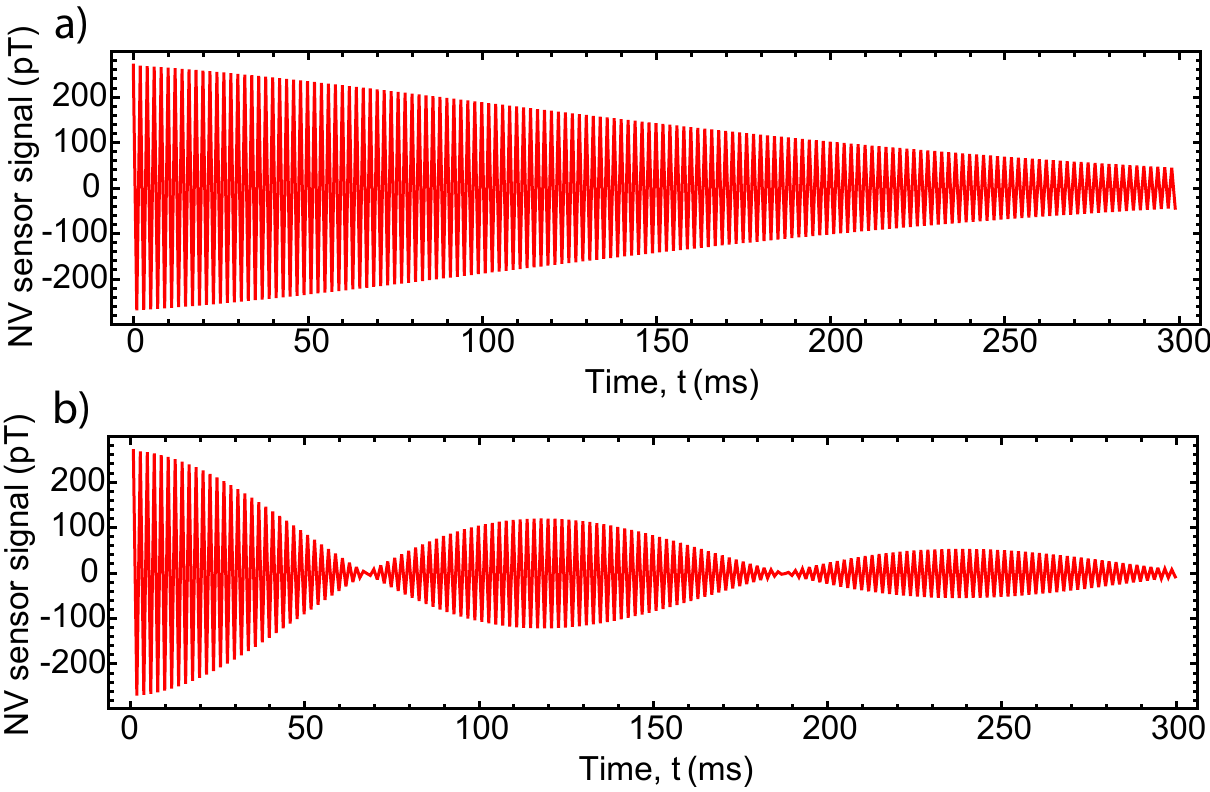}
    \caption{\textbf{Modulated $M_z$ simulation.} (a) Spin dynamics simulation of $B_{\rm nuc}(t)$ of water under the experimental conditions of Fig.~\ref{fig:fig1p5} of the main text: $B_0=0.32~{\rm T}$, Ramsey delay $\tau=0$, and nuclear $T_1=0.6~{\rm s}$. Here we assume hard RF $\pi$ pulses with a frequency detuning error of $1.5~{\rm ppm}$ and an amplitude error of $2.5\%$. (b) Same as (a), but with a frequency detuning error of $6~{\rm ppm}$. }
    \label{fig:sup_water_3kg_beats}
\end{figure}

In our experiments, the effect of RF pulse errors was not too severe, as the absolute RF pulse detuning from any given NMR peak typically did not exceed $30~{\rm Hz}$. As seen in Fig.~\ref{fig:fig1p5}, when driving water (single NMR frequency), the modulated $M_z$ signal mostly follows the ideal behavior, with a decay time $T_{\rm meas}\approx0.2~{\rm s}$ that is not far from the nuclear $T_1\approx0.6~{\rm s}$ limit. To validate the general behavior, we conducted spin dynamics simulations using the Spinach Matlab package \cite{HOG2011}. 

Figure~\ref{fig:sup_water_3kg_beats}(a) shows an example simulation of the modulated $M_z$ signal, $B_{\rm nuc}(t)$, of water under the experimental conditions of Fig.~\ref{fig:fig1p5}--$B_0=0.32~{\rm T}$, Ramsey delay $\tau=0$, and nuclear $T_1=0.6~{\rm s}$. We assume hard RF $\pi$ pulses with a frequency error of $1.5~{\rm ppm}$ (corresponding to $21~{\rm Hz}$) and an amplitude error of $2.5\%$ (see Table~\ref{tab:paramtab}). With these choices, we find a simple monotonically-decaying response that approximately matches the experimental results of Fig.~\ref{fig:fig1p5}. 

\begin{figure}[hbt]
    \centering    \includegraphics[width=\columnwidth]{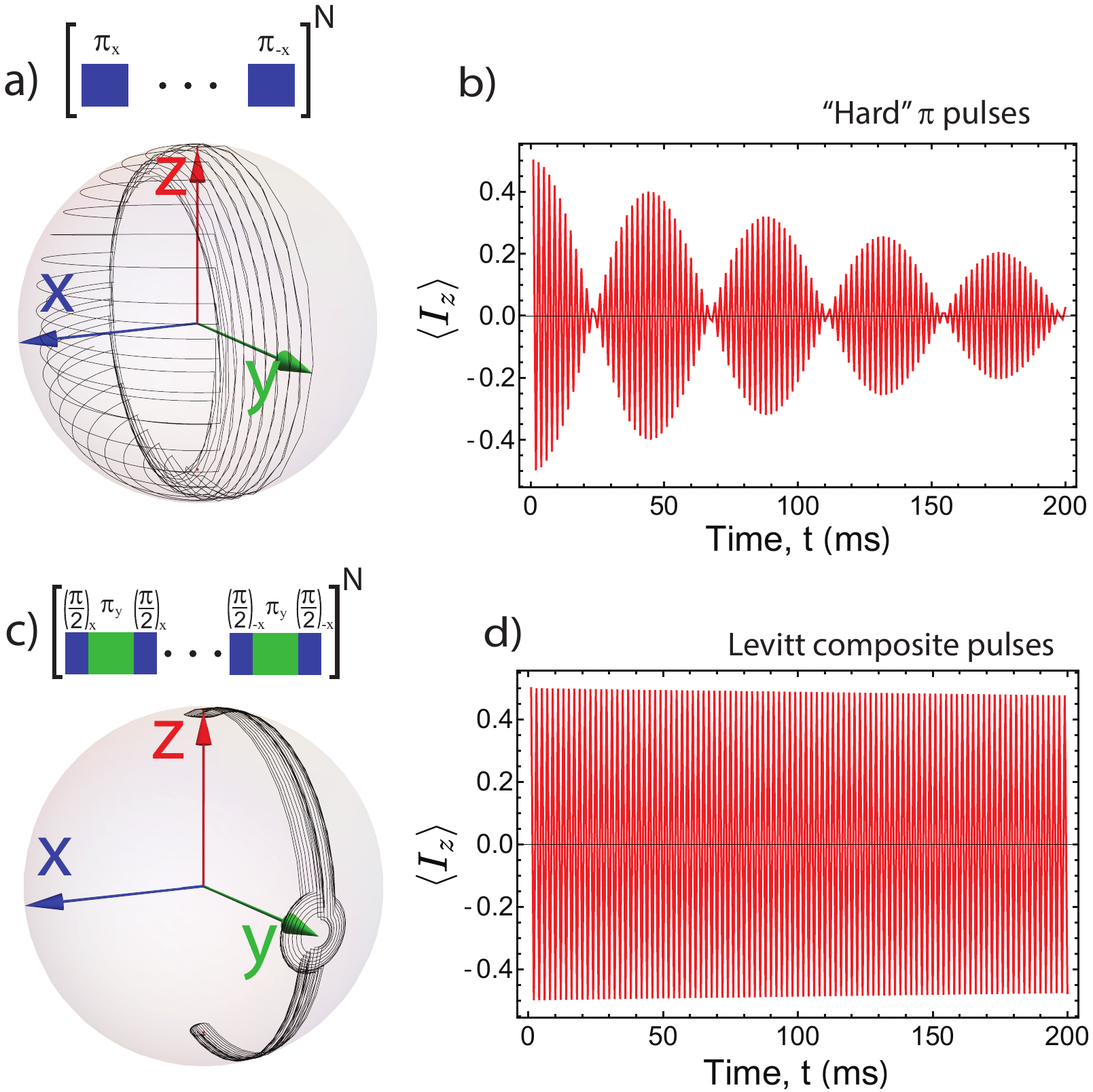}
    \caption{\textbf{Magnetization trajectories.} (a) Nuclear spin vector trajectories under a series of hard, phase-cycled RF pulses. A detuning error of 6 ppm at 1.5 T and an amplitude error of $2.5\%$ is present in both cases. bc) Simulated proton $\langle I_z\rangle(t)$ using the conditions of (a). (c) Same as (a) except using phase-cycled Levitt-Freeman composite inversion pulses of $4~{\rm \upmu s}$ duration separated by $1~{\rm ms}$. (d) Same as (b) except with the composite pulses in (c). In (b,c), we assume unity spin polarization. 
 }
    \label{fig_sup:Bloch_supp}
\end{figure}

To investigate the effects of pulse errors further, we performed a second simulation with the same amplitude error, but now set the detuning error to $6~{\rm ppm}~(84~{\rm Hz})$. The resulting simulated $B_{\rm nuc}(t)$ is shown in Fig.~\ref{fig:sup_water_3kg_beats}(b). Beat notes are observed, which would not be present if the RF pulses produced ideal nuclear-spin inversions. While the $6~{\rm ppm}$ frequency error is larger than that required in our experiments (i.e. chemical shifts are $\lesssim2.5~{\rm ppm}$ for ethanol), the behavior is undesirable and an important consideration for future development efforts. We chose to simulate a $6~{\rm ppm}$ frequency detuning error in Fig.~\ref{fig:fig4_future}(b) to highlight this effect. The presence of beat notes in the $B_{\rm nuc}(t)$ signals indicates that only a short interval can be averaged when computing $\overline{B_{\rm nuc}}$, effectively reducing the sequence duty cycle, $\delta$. This is a major reason why the hard-pulses Ramsey-$M_z$ curve in Fig.~\ref{fig:fig4_future}(b) of the main text has a significantly lower expected weighted signal, $\overline{B_{\rm nuc}}\sqrt{\delta}/\rho$, than for conventional, inductive $M_{\perp}$ detection. It is also related to why the hard-pulses Ramsey-$M_z$ curve in Fig.~\ref{fig:fig4_future}(b) has a non-monotonic response at $B_0\gtrsim2~{\rm T}$.

\begin{table*}[hbt]
    \centering
    \begin{tabular}{| c | c | c | c | c | c | c |}
    \hline
       \textbf{\,Simulation\,} & \textbf{\,$\pi$ pulse length\,} & $~~\mathbf{T_1}~~$ &$~~\mathbf{T_2^{\ast}}~~$ & \textbf{~Frequency error~} & \textbf{~Amplitude error~} & $~\mathbf{B_0}~$\\ 
       \hline
        Fig.~\ref{fig:fig4_future}(a) & $2~{\rm \upmu s}$& $5~{\rm s}$ & $0.1~{\rm s}$ & Resonant with ${\rm CH}_2$& $\mu{=}3 \%,\sigma{=}5\%$ & 1
         \\ \hline
         Fig.~\ref{fig:fig4_future}(b) & $2~{\rm \upmu s}$ & $5~{\rm s}$ & $0.1~{\rm s}$ & $6~{\rm ppm}$ & $\mu{=}3\%,\sigma{=}5\%$ & $0.1{\mbox{-}}3~{\rm T}$ \\
         \hline
         Fig.~\ref{fig:sup_water_3kg_beats}(a) & $20~{\rm \upmu s}$ & $0.6~{\rm s}$ & $0.1~{\rm s}$ & $1.5~{\rm ppm}$ & $2.5\%$ & $0.32~{\rm T}$ \\
         \hline Fig.~\ref{fig:sup_water_3kg_beats}(b) & $20~{\rm \upmu s}$ & $0.6~{\rm s}$ & $0.1~{\rm s}$ & $6~{\rm ppm}$ & $2.5\%$ & $0.32~{\rm T}$ \\
         \hline
         Fig.~\ref{fig_sup:Bloch_supp}(b,d) & $2~{\rm \upmu s}$ & $5~{\rm s}$ & $0.1~{\rm s}$ & $6~{\rm ppm}$ & $2.5\%$ & $1.5~{\rm T}$ \\
         \hline
         \end{tabular}
    \caption{\textbf{Simulation parameters.} If an expectation value $\mu$ and a standard deviation $\sigma$ are provided, then the amplitude errors are described by the Gaussian distribution of Eq.~\eqref{eq:si_normdist}. When composite Levitt-Freeman inversion pulses are used, the total inversion pulse length is twice the $\pi$ pulse length.
    }
    \label{tab:paramtab}
\end{table*}

To better understand and visualize the impact of RF pulse errors, we conducted simulations at a higher magnetic field, $B_0=1.5~{\rm T}$ where the absolute frequency errors can be substantial. Figure~\ref{fig_sup:Bloch_supp}(a) shows the Bloch-sphere spin trajectory of protons in water ($T_1=5~{\rm s}$, $T_2^{\ast}=100~{\rm ms}$) under a series of hard RF inversion pulses separated by $1~{\rm ms}$ intervals of free evolution. The simulation assumes an RF amplitude error of $2.5\%$ and a frequency detuning error of $6~{\rm ppm}$ which now corresponds to an absolute error of $380~{\rm Hz}$ (Table~\ref{tab:paramtab}). With ideal inversion pulses, the spin vector would simply oscillate along a great circle of the Bloch sphere, going back and forth between the two poles. However, under these conditions for pulse errors, the spin vector rapidly deviates from the intended trajectory, tracing out a complicated path of the Bloch sphere, a considerable fraction of which is near the equator. Figure~\ref{fig_sup:Bloch_supp}(b) plots the longitudinal component of the magnetization $\langle I_z \rangle$ as a function of time, $t$. Anti-nodes in the beat pattern correspond to times when the spin vector lies along the equator. Each nodal revival has the opposite phase of the previous oscillation, leading to a partial cancellation when averaging over longer intervals. 

Fortunately, there is a simple remedy for this behavior long known in the NMR literature. Specifically, we propose to replace the hard RF pulses with composite phase-cycled Levitt-Freeman composite inversion pulses ($90_x-180_y-90_x$)~\cite{LEV1979,TYC1983}. Figure~\ref{fig_sup:Bloch_supp}(c) shows the trajectory of the spin vector using Levitt-Freeman inversion pulses. The spin vector follows the ideal case much more closely. Figure~\ref{fig_sup:Bloch_supp}(d) shows the corresponding $\langle I_z\rangle(t)$ plot. The only observed deviation from ideal behavior is a slight decay due to spin relaxation. This leads to an increased weighted signal, $\overline{B_{\rm nuc}}\sqrt{\delta}/\rho$, compared to "hard" pulses, as seen in Fig.~\ref{fig:fig4_future}(b) of the main text. The interval over which $B_{\rm nuc}(t)$ is averaged is longer, resulting in a higher duty cycle. This is due to a combination of the absence of beat notes and anti-phased revivals as well as a slower overall decay. The latter is due to the spin vector spending most of the time near the poles of the Bloch sphere, where only the longitudinal relaxation, $T_1>T_2^{\ast}$, contributes to relaxation.

\subsection{Simulation parameters}
\label{sec:SI_simulationparam}

Table~\ref{tab:paramtab} lists the simulations parameters used to generate curves in Figs.~\ref{fig:fig4_future}, \ref{fig:sup_water_3kg_beats}, and \ref{fig_sup:Bloch_supp}. The amplitude error is given as a percent error of an ideal, resonant $\pi$ pulse amplitude, $B_{10}$. For Fig.~\ref{fig:fig4_future}, we assume a distribution of amplitude errors. Specifically, we define an amplitude distribution $B_1=B_{10}(1+\Delta B_1)$, where $\Delta B_1$ is normally distributed:
\begin{equation}
\label{eq:si_normdist}
P(\Delta B_1)=\frac{1}{\sqrt{2\pi \sigma^2} }e^{\frac{\left(\Delta B_1 - \mu\right)^2}{2\sigma^2}}.
\end{equation}
Spin dynamics simulation were run at $5$ different values $\Delta B_1 = [-0.05, -0.025, 0, 0.025, 0.05]$. The simulations were weighted by $P(\Delta B_1)$ and added together. 

In Fig.~\ref{fig:fig4_future}(a), we simulated the spectrum of ethanol at $B_0=1.5~{\rm T}$ to explore how RF pulse errors might impact the Ramsey-$M_z$ NMR spectrum of a molecule with both chemical shifts and $J$-couplings. The simulation was done for $100$ different Ramsey delay values uniformly sampled from $\tau=0$ to $\tau=150~{\rm ms}$. For each $\tau$ value, a $B_{\rm nuc}(t)$ curve was simulated from $t=0$ to $t=0.75~{\rm s}$, with RF inversion pulses occurring every $1~{\rm ms}$. The simulated data were then processed similarly to the experimental data (\ref{SI:sig_proc}), with the only difference being the absence of a windowing function.

To generate the curves shown in Fig.~\ref{fig:fig4_future}(b) of the main text, we simulated the spin trajectories under the "hard" $\pi$ pulses and the composite Levitt-Freeman inversion pulses out to $T_{\rm acq}=5~{\rm s}$. To find the optimum averaging interval used to compute $\overline{B_{\rm nuc}}$, we selected the value that maximizes the weighted signal amplitude $\overline{B_{\rm nuc}}\sqrt{\delta}/\rho$. Here, $\overline{B_{\rm nuc}}$ refers to the $\tau=0$ maximum signal. To convert $\langle I_z\rangle(t)$ to $B_{\rm nuc}(t)$, we use the magnetostatic calculations described in~\ref{sec:SI_nucmag}. For the inductive $M_{\perp}$ curve, we assume a $1~{\rm nL}$ cylindrical analyte volume with height$=$diameter$=h$, corresponding to a analyte-volume-averaged-field of $B_{\rm nuc,0}=865~{\rm pT}$ at $B_0=0.32~{\rm T}$. For NV-detected Ramsey-$M_z$ curves, we use the analyte-volume-averaged-field of $B_{\rm nuc,0}=340~{\rm pT}$. In both cases, we then scale the nuclear field linearly with $B_0$. For the Ramsey-$M_z$ simulation, we assumed a (111)-cut diamond geometry, which gives a two-fold larger $B_{\rm nuc}$ than the (110)-cut case used in experiments~(\ref{SI:diamond_cut}). Moreover, we assume the optimal NV illumination volume dimensions of diameter$=h$, height$=h/3$ (\ref{SI:opt_dim}).

\section{Ramsey-ENDOR}

\label{SI:Ramsey_ENDOR}

\begin{figure*}[ht]
    \centering
    \includegraphics[width=0.99\linewidth]{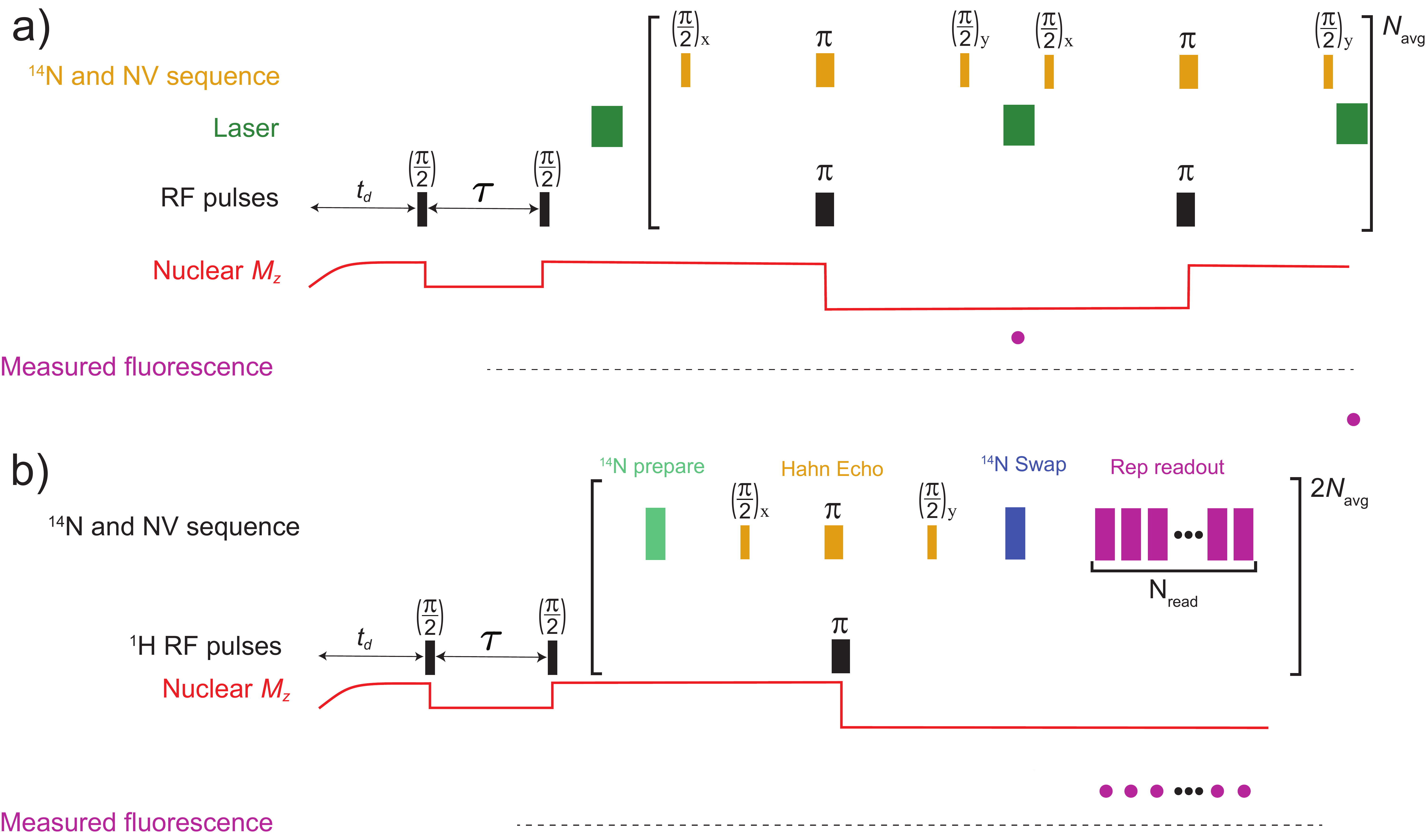}
    \caption{\textbf{Ramsey-ENDOR pulse sequence.} (a) The Ramsey-ENDOR sequence with synchronized NV Hahn-Echo and nuclear inversion pulses. (b) A $^{14}\rm N$ quantum memory enhanced Ramsey-ENDOR sequence. }
    \label{fig:SI_Ramdoor}
\end{figure*}

Our diamond magnetometer used a CW-ODMR protocol to measure nuclear magnetization. This approach is relatively easy to implement, doesn't require high microwave or laser power, is broadband (allowing for a wide range of modulated $M_z$ frequencies), and it has an advantage that signals during the time when RF pulses are applied can simply be removed. However, the sensitivity is limited by the NV electron-spin dephasing time ($T_2^{\ast}\approx1~{\rm \upmu s}$), which results in a relatively poor sensitivity. 

As discussed in the main text, the highest-sensitivity diamond magnetometer protocols involve pulsed ODMR measurements of kHz-MHz magnetic fields~\cite{TAY2008}. These protocols involve electron-spin microwave pulse sequences, such as Hahn echo, that refocus electron spin dephasing. The sensitivity is then limited only by the much longer $T_2\gtrsim100~{\rm \upmu s}$ transverse decoherence time. 

In a Hahn-echo type AC measurement, the photoelectron shot noise limited sensitivity is given by:
\begin{equation}
    \eta_{\rm psn} = \frac{1}{ 2 \pi\, \gamma_{nv} \, C_{\rm eff} \sqrt{\, \xi \, n_{\rm nv} \, V_{\rm ill} T_2}},
\end{equation}
where $C_{\rm eff}$ is the effective fluorescence contrast, $n_{\rm nv}$ is the NV number density, $\xi$ is the photoelectron collection efficiency (number of photoelectrons detected per NV center per readout), and $V_{\rm ill}$ is the NV sensor volume, and we have assumed a square-waveform magnetic field of frequency ${\sim}1/T_2$ is being detected. The product $T_2 n_{\rm nv}$ is constrained by NV-NV dipole-dipole interactions~\cite{TAY2008,ACO2009,BAU2018,EIC2019}, but diamonds with $n_{\rm NV} \approx 6 \times 10^{16} ~{\rm cm^{-3}}$ and $T_2\approx200~{\rm \upmu s}$ are routinely available. In an optimized optical setup~\cite{DUA2019}, exploiting the polarization-selective absorption and emission of NV centers~\cite{ALEG2007}, $C_{\rm eff} \approx 2\%$, and $\xi \approx 10\%$ are plausible estimates. These assumptions arrive at a Hahn-Echo AC sensitivity of $\eta_{\rm psn} \approx 0.5~{\rm pT \,s^{1/2}}$ for a $30~{\rm \upmu m}$ thick diamond illuminated by a $100~{\rm \upmu m}$ diameter laser beam. 

Figure~\ref{fig:SI_Ramdoor}(a) shows a proposed protocol, ``Ramsey-ENDOR'', that could be used for $M_z$-detected NMR with a pulsed AC diamond magnetometer. As in the Ramsey-$M_z$ protocol, an RF nuclear Ramsey sequence encodes the phase accumulated by analyte nuclear magnetization during a free precession interval, $\tau$, into an $M_z$ amplitude. In order to modulate and sensitively detect the $M_z(\tau)$ signal, RF nuclear-spin inversion pulses are embedded within electron-spin microwave Hahn-echo sequences. The RF inversion pulses may be applied at the same time as the NV $\pi$ pulses or slightly before or after to avoid any possible technical issues with temporal overlap. The RF inversion pulses can be simple hard pulses or optimized composite pulses, as needed. In either case they convert $M_z$ into a square wave that is sensitively detected by the NV hahn echo sequence. 

Figure~\ref{fig:SI_Ramdoor}(b) shows a potential upgrade to the Ramsey-ENDOR protocol based on repetitive readout of the $^{14}$N nuclear spins intrinsic to each NV center~\cite{PAG2014,ARU2023}. In this case the Hahn echo sequences are modified as follows. First, the $^{14}$N are polarized into $m_i=0$ using a sequence of: i) a laser pulse that polarizes NV centers into $m_s=0$, ii) a selective microwave $\pi$ pulse that drives NV centers into $m_s=\pm1$ only if nuclear spins are in $m_i=0$, and iii) a double-quantum RF pulse that drives nuclear spins from $m_i=\pm1$ to $m_i=0$ within the $m_s=0$ manifold. Next, the microwave Hahn echo sequence (with synchronized embedded RF inversion pulse) proceeds in the same manner as the normal Ramsey-ENDOR sequence in Fig.~\ref{fig:SI_Ramdoor}(a). Subsequently, the NV $m_s$ state is ``swapped'' with the $^{14}$N $m_i$ state by applying an RF pulse that selectively drives nuclear spins to $m_i=\pm1$ only within the $m_s=0$ manifold. A laser pulse is applied to polarize NV centers into $m_s=0$ and then a repetitive readout is applied consisting of: i) a selective microwave $\pi$ pulse that drives NV centers from $m_s=0$ to $m_s=\pm1$ only if the nuclear spin is in $m_i=0$ and ii) a laser pulse which reads out the NV population in $m_s=0$ and then repolarizes NV centers. We expect ${\sim}50$ repetitive readout sequences of ${\sim}4~{\rm \upmu s}$ duration can be applied such that the measurement time is only increased by a factor of ${\sim}2$. 

We project that this approach could improve the magnetometer sensitivity by a factor of ${\sim}5$ to $\eta_{\rm psn}\approx 0.1~{\rm pT\,s^{1/2}}$. Note that the impact of microwave phase noise would also need to be mitigated to this level. This may be accomplished by using two-point gradiometry as demonstrated in Ref.~\cite{BER2024}. With these improvements, we estimate a minimum detectable proton concentration sensitivity (SNR=3) of $3\,\eta_{\rm psn}/(\overline{B_{\rm nuc}}\sqrt{\delta}/\rho)\approx40~{\rm mM\,s^{1/2}}$ is possible, for $0.7~{\rm nL}$ of analyte at $B_0=3~{\rm T}$. Here, we assumed the simulated value of $\overline{B_{\rm nuc}}\sqrt{\delta}/\rho\approx7~{\rm pT/M}$ for the weighted signal strength, as shown in Fig.~\ref{fig:fig4_future}(b) of the main text, and an analyte cylinder volume of diameter $100~{\rm \upmu m}$ and height of $95~{\rm \upmu m}$. This is an improvement of ${\sim}100$ over the best concentration threshold reported for sub-nL microcoil NMR in the literature. For example Ref.~\cite{GRI2017} reported a ``mass'' sensitivity of $1900~{\rm pmol}$ in a ${\sim}0.5~{\rm nL}$ volume, which corresponds to a minimum detectable proton concentration of ${\sim}4~{\rm M\,s^{1/2}}$. Note the minimum detectable proton concentration assumes a peak degeneracy of 1 (no protons in the molecule are magnetically equivalent). In cases of higher degeneracy, such as water (degeneracy$=2$), the minimum detectable proton concentration should be divided by the degeneracy.

\clearpage
% \bibliography{references2}

%

\end{document}